\documentclass[lettersize,journal]{IEEEtran}
\usepackage{amsmath,amsthm,amssymb,amsfonts}
\usepackage{cuted}
\usepackage{algorithmic}
\usepackage{algorithm}
\usepackage{array}
\usepackage{textcomp}
\usepackage{stfloats}
\usepackage{url}
\usepackage{verbatim}
\usepackage{graphicx}
\usepackage{subcaption}
\usepackage{multirow}
\usepackage{cite}
\usepackage{stmaryrd}
\usepackage{color}
\usepackage{xcolor}
\usepackage{etoolbox}
\newtoggle{comments}
\newtoggle{anon}
\toggletrue{comments}
\toggletrue{anon}
\usepackage{todonotes}
\usepackage{makecell}
\usepackage{booktabs}

\iftoggle{comments}{%
  \newcommand{\melody}[2][]{\todo[color=blue!30,#1]{\textsf{Melody:} #2}}
  \newcommand{\jiabo}[2][]{\todo[color=yellow!30,#1]{\textsf{Jiabo:} #2}}
  \newcommand{\ivan}[2][]{\todo[color=red!30,#1]{\textsf{Ivan:} #2}}
}{%
  \newcommand{\melody}[2][]{}
  \newcommand{\jiabo}[2][]{}
  \newcommand{\ivan}[2][]{}  
}

\newtheorem{lemma}{Lemma}[section]

\newtheorem{claim}[lemma]{Claim}
\newtheorem{prop}[lemma]{Proposition}

\newtheorem{defn}{Definition}

\theoremstyle{remark}
\newtheorem{rmk}{Remark}
\begin{document}

\title{A Failure-Free and Efficient Discrete Laplace Distribution for Differential Privacy in MPC}

\author{Ivan Tjuawinata$^\dag$, Jiabo Wang$^\dag$, Mengmeng Yang, Shanxiang Lyu, Huaxiong Wang, Kwok-Yan Lam
\thanks{J. Wang, I. Tjuawinata are with the Strategic Centre for
Research in Privacy-Preserving Technologies $\&$ Systems, Nanyang Technological
University, Singapore. M. Yang is with Information Security and Privacy Group, Data61, CSIRO, Australia.
Shanxiang Lyu is with the College of Cyber Security, Jinan University, Guangzhou, China.
Huaxiong Wang is with the Division of Mathematical Sciences, Nanyang Technological University, Singapore. Kwok-Yan Lam is with the College of Computing and Data Science, Nanyang Technological University, Singapore.
}
\thanks{$^\dag$ The first two authors have the same amount of contributions.}
\thanks{jiabo.wang@ntu.edu.sg}}

\markboth{Journal of \LaTeX\ Class Files,~Vol.~14, No.~8, August~2021}%
{Shell \MakeLowercase{\textit{et al.}}: A Sample Article Using IEEEtran.cls for IEEE Journals}


\maketitle

\begin{abstract}

 In an MPC-protected distributed computation, although the use of MPC assures data privacy during computation, sensitive information may still be inferred by curious MPC participants from the computation output. This can be observed, for instance, in the inference attacks on either federated learning or a more standard statistical computation with distributed inputs. In this work, we address this output privacy issue by proposing a discrete and bounded Laplace-inspired perturbation mechanism along with a secure realization of this mechanism using MPC. The proposed mechanism strictly adheres to a zero failure probability, overcoming the limitation encountered on other existing bounded and discrete variants of Laplace perturbation. We provide analyses of the proposed differential privacy (DP) perturbation in terms of its privacy and utility. Additionally, we designed MPC protocols to implement this mechanism and presented performance benchmarks based on our experimental setup.  The MPC realization of the proposed mechanism exhibits a complexity similar to the state-of-the-art discrete Gaussian mechanism, which can be considered an alternative with comparable efficiency while providing stronger differential privacy guarantee. Moreover, efficiency of the proposed scheme can be further enhanced by performing the noise generation offline while leaving the perturbation phase online.

\end{abstract}

\begin{IEEEkeywords}
Differential Privacy, Secure Multi-Party Computation, Discrete Laplace Distribution, Failure Probability
\end{IEEEkeywords}

\section{Introduction}
\IEEEPARstart{D}{ifferential} privacy (DP) is an efficient and lightweight approach in the domain of privacy-preserving technologies, particularly when compared to cryptographic methodologies. 
The objective of DP is to protect private information that may be reconstructed from the output by introducing some noise to the output of such calculation. Such protection is provided while also maintaining the utility of the perturbed output. Due to how DP provides the privacy guarantee, it is especially useful in the scenario of trend learning or statistical calculation where exact output may not be necessary.  DP mechanisms, depending on trust model, are classified into two types of models, i.e. local DP and central DP. In the former model, individuals perturb their data locally   before it being sent to the server for the function calculation. This is done to provide privacy protection against a semi-honest server who performs the calculation. On the other hand, the latter model usually relies on the availability of a trusted aggregator  that collects the input data from the data owners in the clear, performs the necessary calculation, and perturbs the computation result to provide privacy protection against the querier.

In this work, we focus on a scenario where private data is distributed and collaborative computation is needed to gain insight of the data while individuals' data privacy is to be preserved. Real-world use cases of this kind include machine learning model training using distributed and private data (e.g. federated learning (FL), SecureML \cite{fl06,secureML17}) and secure analytics on distributed databases \cite{bater2017smcql}.  In the case of federated learning, while FL itself effectively mitigates the privacy risk and facilitates model training on distributed private data, it inevitably faces with two types of privacy issues from the aggregator. Firstly, the aggregator, usually a third-party server, can infer individuals' private data given their updates amid the FL procedures \cite{HAP17}. Secondly, even with a shuffler between clients and the server, once the aggregation is done, the malicious central server can reconstruct private training data given the model coefficients, for instance, using the technique of ``deep leakage from gradient'' \cite{NEURIPS2019}. Perturbing the updates locally at each round on clients' side in FL helps preserve the privacy, but it comes at the expense of accuracy.

To address the above issues, researchers applied the \textit{distributed DP mechanism} where the clients slightly perturb their private data locally such that the aggregated perturbation meets the DP guarantee \cite{pillutla2022differentially,KLS22}. Another line of research resorts to cryptographic tools. The SecAgg \cite{secagg17,secagg20} was proposed such that the central server, as the aggregator, can no longer see individual's updates amid each round of iteration. Instead, it can only learn the aggregate of such updates. Nonetheless, once FL is done and the updated model is derived, the central server or a third party having access to the updated model still have the opportunity to perform inference attack and reverse engineering to infer information about the private training data.

In the case of secure analytics, especially for those with distributed databases facilitated by secure multiparty computation (MPC), the cryptographic tools only  provide security during the computation process. Once the answer is given to a curious querier, there is always a likelihood of information leakage which can be inferred from the computation result. Furthermore, a malicious querier may strategically devise a series of queries such that the private data can be recovered from the query results. Although local DP solutions can ensure privacy guarantees in this scenario, they come with significant limitations in terms of utility due to the large amount of noise introduced during the computation process. In some cases, this noise can make the data nearly useless or even lead to faulty conclusions \cite{ZWW23}. Therefore, it may not be the ideal choice for scenarios that involve various types of computations. It is then natural to instead consider the use of a central DP solution. Here to resolve the issue of the need of a trusted server, it can be realized by multiparty computation. By using central DP solution, calculation can be done using the original input to provide the correct analytics result and perturbation may be done to the result, for instance, while it is still being secretly shared  \cite{Cat18,keller2024secure,WYF23}. Examples of such solutions are the secure distributed differentially private computation of median and heavy hitters which can be found in \cite{BK20,BK21}.

Another use case in the category of secure analytics on distributed data is private set operations, such as private set intersection (PSI) with cardinality and PSI-sum. 
In order to provide privacy guarantee throughout the calculation, a generic MPC-based realization of the private set operation framework called OPPRF PSI, which is also known as circuit-based PSI \cite{pinkas2019efficient} , may be performed to obtain the output as secret shares.  However, given the cleartext output of those secure computation, existing inference attacks \cite{kacsmar2020differentially,jiang2023anonpsi} may be used to violate the data privacy. In order to provide output privacy in such realization, a DP mechanism that is compatible with the setting of the protocol and can also be jointly realized using MPC is desired.

To achieve differentially private addition-noise perturbation mechanisms that are compatible with MPC, we need to design a variant of such additive noise when the input and output space are discrete, bounded, and with finite precision while maintaining the differential privacy guarantee \cite{canonne2020discrete}.  
Additive-noise-based perturbation mechanisms such as Laplace and Gaussian are originally defined over the real number $\mathbb{R}$ or real-valued vectors $\mathbb{R}^n$ which are both continuous and unbounded. Because of this, they are not compatible with either the digital computing devices, that only have finite data precision, or the setting of MPC which requires any value to be encoded to a finite field or finite ring element. This necessitates the step of discretizing and truncating the additive-noise-based perturbation mechanisms to transform their output to a discrete and bounded space. However, such transformation needs to be done carefully. A direct discretization and truncation, for example, using the floating-point encoding to the original output will affect the privacy guarantee and, indeed, there have been existing attacks on such transformation \cite{mironov2012significance,Jin2021AreWT}. In order to prevent such attacks, transformation needs to be done carefully with special attention given to its impact on privacy guarantee \cite{KLS22,canonne2020discrete}.

In most work exploring the use of DP in an MPC setting, discrete Gaussian distribution has been the primary perturbation mechanism being considered. The tighter tails of a Gaussian distribution makes it better suited for applications with a high degree of composition, as it introduces relatively smaller noise compared to a Laplace distribution while providing a similar privacy guarantee. \cite{WYF23,canonne2020discrete}. However, such approach comes with some limitations. Firstly, we note that even in its standard variant defined over real numbers, the privacy guarantee provided by Gaussian perturbation has a non-zero failure probability \cite{keller2024secure,canonne2020discrete,KLS22}. In order to achieve zero failure probability, Laplace perturbation may be a better candidate for the base of the discrete and bounded perturbation. However, despite the standard Laplace distribution over real numbers achieves $(\epsilon, 0)$-DP guarantee, the same cannot be said in any existing work on discrete Laplace distribution \cite{pentyala2022training}. Secondly, efficiently generating discrete Gaussian samples has been a bottleneck in some digital signature schemes  \cite{wang2023polar,micciancio2017gaussian}. The problem becomes even more severe when distributed scenario through MPC is considered. Such limitations provide us with two questions:
\begin{enumerate}
    \item Is it possible to design a discrete and bounded Laplace-inspired perturbation that provides differential privacy guarantee with a zero failure probability?
    \item Is it possible for such perturbation mechanism to be MPC friendly with a low complexity requirement?
\end{enumerate}    

In this work, we show that the two questions posed above have affirmative answers where we provide the designs satisfying such requirements. In the following subsection, we discuss our contributions in more details.

\subsection{Contributions}
The contributions of this work are as follows.
\begin{itemize}
    \item 
    We developed two $\epsilon-$DP mechanisms with two types of bounded and discrete perturbations. The perturbations follow the truncated discrete Laplace distribution and the truncated cumulative Laplace distribution, respectively. Compared with existing discretized perturbations, the proposed ones are remarked by a failure-free design.
    \item Secondly, we analyzed the privacy guarantee of the proposed DP mechanisms in the context of generic and distance-based DP, respectively. Additionally, we also analyzed the utility of the proposed DP mechanisms. 
    \item Thirdly, we proposed an MPC realization of the proposed DP mechanisms.  We further note that our realization allows the noise generation to be done during the offline phase while only a much more efficient noise injection mechanism needs to be performed online. This further improves the amortized complexity of our proposed protocol. Benchmarks based on experimental setup are also given.
\end{itemize}

\section{Preliminaries}
\subsection{Differential Privacy}
Differential privacy was proposed by Dwork \textit{et al.} in \cite{Dwork06}. A differentially private randomized mechanism is said to protect data privacy in the sense that the randomized mechanism behaves similarly on neighboring data sets, two data sets with one differences, either through the removal, addition or replacement of one data record in the dataset, i.e., $D,D'$ is said to be neighbouring if $|(D\setminus D')\cup(D'\setminus D)|=1.$  
\begin{defn}[Differential Privacy]
    Let $\mathcal{X}$ be the data space and $\mathcal{D}\subseteq \mathbb{P}(\mathcal{X})$ be the data set space containing subsets of $\mathcal{X}.$ We further let $\mathcal{R}$ be the output space containing the possible perturbed response expected.
    A randomized algorithm $\mathcal{M}:\mathcal{D}\rightarrow \mathcal{R}$ is $(\epsilon,\delta)-$differentially private if for all subset ${S}\subseteq\mathcal{R}$ and for all neighboring data set $D$ and $D'\in\mathcal{D}$, we have
    \[
        Pr[\mathcal{M}(D)\in S] \leq e^{\epsilon}Pr[\mathcal{M}(D')\in S]+\delta.
    \]
    If $\delta=0$, we say that $\mathcal{M}$ is $\epsilon-$differentially private.
\end{defn}

Laplace mechanism is one of the first proposed perturbation mechanism \cite{dwork2006calibrating} that satisfies the differential privacy definition.

\begin{defn}[Laplace Mechanism \cite{dwork2006calibrating}]
    Let $\mathcal{X}$ be a data space and $\mathcal{D}\subseteq\mathbb{P}(\mathcal{X})$ be the data set space. 
    Suppose further that there is a statistical calculation $f:\mathcal{D}\rightarrow \mathbb{R}$ which calculates some statistical calculation based on a dataset $D\in \mathcal{D},$ outputting a real number $f(D).$ Here let $\Delta_1$ be the $\ell_1$-sensitivity be the sensitivity of $f.$ That is, $\Delta_1 = \max_{D,D'\in \mathcal{D}: |(D\setminus D')\cup (D'\setminus D)|\leq 1}|f(D)-f(D')|.$
    
    A Laplace perturbation mechanism $\mathcal{M}_f:\mathcal{D}\rightarrow \mathbb{R}$ with respect to $f$ maps $D\in \mathcal{D}$ to $\mathcal{M}_f(D) = f(D) +r$ where $r$ follows the Laplace distribution with mean $0.$ That is, for Laplace distribution with parameter $(0,\sigma),$ the density function $g_L$ of $\mathbf{r}$ is  $g_L(\mathbf{r})=\frac{1}{2\sigma}e^{-\frac{|r|}{\sigma}}$. It is shown that for any $\epsilon>0,$ when $\sigma = \Delta_1/\epsilon,$ the perturbation mechanism $\mathcal{M}_f$ provides $\epsilon$-differential privacy. 

\end{defn}

In this work, we consider the discrete variant of Laplace distribution. In the following, we consider a straightforward discretization of Laplace distribution.
\begin{defn}[Discrete Laplace Distribution]
    Let $S\subseteq \mathbb{R}$ be a discrete subspace of $\mathbb{R}.$ For parameter $\sigma>0$, the discrete Laplace distribution $\mathcal{L}_{S}(\sigma)$ is defined as
    \[
        f_{\mathcal{L}_{S}}(x;\sigma) = c\cdot e^{-\frac{|x|}{\sigma}}~\textrm{for~}x\in S
    \]
    where $c$ is a normalizing constant, that is, $c=\sum_{x\in S} e^{-\frac{|x|}{\sigma}}$. 

\end{defn}
\subsection{Secure Multi-Party Computation}

In a secure multi-party computation protocol, $n$ parties $P_1,\cdots,P_n$, each having a private input $x_i$ for $i=1,\cdots,n$, jointly and securely compute a function $z=f(x_1,\cdots,x_n)$ in a way such that nothing else about $x_i$ is learned except the result $z$ and what is leaked from $z$ about $x_i$. In an ideal world, this functionality can be fulfilled by a trusted third party who is able to see the private inputs and computes the result. Secure multi-party computation protocols are designed to realize such functionality without the existence of a trusted third party such that it is indistinguishable from the ideal solution \cite{cramer2015secure}.

In a general purpose MPC protocol, we compile the function $f$ into either an arithmetic or an boolean circuit and apply garbled-circuit based protocols or secret-sharing based protocols to securely and correctly compute $f$ \cite{escudero2022introduction}. In addition to generic MPC, Some application-oriented MPC protocols, e.g. Private Set Intersection, leverages comprehensive techniques in the field of cryptography and computer science to facilitate the realization of versatile functionalities. 

In this work, we are especially interested in providing output privacy to the output of a secret-sharing based MPC protocol before the final output is revealed. 

\subsection{Fixed Point Encoding for MPC}

In order to make use of MPC to provide privacy in the computation over real numbers, the values need to first be encoded using a finite precision as field elements where the encoding needs to be compatible with the field operations. In this section, we consider the fixed point encoding $\varphi_q:\mathbb{R}\rightarrow \mathbb{F}_q$ \cite{Cat18}.

We assume the existence of a positive integer $e$ such that for any value $x$ that is considered, $x \in \mathbb{R}$ and $|x|<2^{e-1}.$ We further assume that the participants agree on a precision parameter $p\in \mathbb{Z}_{>0}$. This implies the existence of a set $S_{e,p}\subseteq \mathbb{R}$ where for any value $y\in S_{e,p},$ there exist $d_{-p},\cdots, d_{e-2}\in \{0,1\}$ such that $y=\pm \sum_{i=-p}^{e-2} d_i 2^{i}.$  Hence, given $x\in \mathbb{R},$ it is firstly encoded to $\tilde{x}\in S_{e,p}$ by truncating the least significant bits, namely $\tilde{x} = 2^{-p}\lfloor x\cdot 2^p\rfloor$. Then, we assume that $q$ is an odd prime that is larger than $2^{p+e}$ and let $\mathbb{F}_q=\left\{-\frac{q-1}{2},\cdots, -1,0,1,\cdots, \frac{q-1}{2}\right\}.$ Given $x\in \mathbb{R}, \varphi_q(x)=\bar{x}$ where $\bar{x}\equiv 2^p\tilde{x}\pmod{q}$. Note that for such encoding process, a truncation protocol is required for any invocation of multiplication. More specifically, given $\bar{x}$ and $\bar{y}\in \mathbb{F}_q,$ if $\hat{z} = \bar{x}\cdot \bar{y},$ then $\hat{z}$ has precision of within $2^{-2p}.$ Hence to avoid the need of infinite precision, a truncation protocol is used to convert $\hat{z}$ back to $\bar{z},$ which has precision of within $2^{-p}$. Here we note that for the perturbation purpose, only addition is required where we add the noise to the value we would like to protect.

\section{Problem Statement}
In this work, we consider the problem of the design of Laplace-inspired perturbation mechanism $\mathcal{M}:\mathcal{R}\rightarrow \mathcal{R}'$ when $\mathcal{R}\subseteq \mathcal{R}'\subseteq\mathbb{R}$  are discrete and bounded subsets of $\mathbb{R}.$ Here for simplicity, we assume that the original value and the perturbed value belong to $S_{e,p}$ and $S_{e+\ell,p}$ respectively for some predefined values $e,\ell,$ and $p.$ Here, we require that such perturbation mechanism provides $(\epsilon,0)\text{-}$differential privacy.

By the use of encoding function described above, if such mechanism exists, it can be directly transformed to a finite field perturbation mechanism. For such transformed mechanism, we are also interested in its realization, especially in a multiparty setting. In such case, such distributed perturbation mechanism can be used to perturb a secretly-shared output before it being revealed to limit the possible privacy leakage through the output.

Lastly, we would also like to require that such distributed perturbation mechanism can be divided to two steps, the noise generation and noise injection mechanisms where noise generation can be done offline and contains the more complex calculation while the online noise injection mechanism is more efficient to provide better amortized complexity.

\section{Truncated Laplace Mechanism}\label{sec:CircPSISol-TruncLapGaus}
Recall that we would like to change two features regarding the space of Laplace distribution, namely, we would like the input and output space be discrete and bounded. In this section, we first consider the modification needed to bound the input and output. We also provide analysis on its privacy and utility. Having the design proposed in this section, it is further modified in the following sections to transform its input and output space to a discrete space.

\subsection{Truncation on Laplace Mechanism}
Now, we consider the truncated Laplace mechanism which will be analyzed in a similar manner as the bounded perturbation mechanism proposed in \cite{YTL22}. Assume the existence of parameters $E=2^{e-1}$ and $L$ such that the considered data $x\in \mathbb{R}$ satisfies $|x|<E$ and for any possible output of the perturbation mechanism $y\in \mathbb{R},$ we have $|y|<E+L.$ Here the parameter $L$ is given such that the distribution of the noise $r$ depends on its size when it is smaller than $L$ while any noise that is larger than $L$ happens uniformly. Given a parameter $\sigma,$ the truncated Laplace mechanism works as follows.
\begin{itemize}
    \item Truncated Laplace mechanism $\mathcal{M}^{(\mathtt{Lap})}_{L,E,\sigma}$ : given a data $x\in[-E,E],$ it is perturbed to $y$ where for $y\in [-L-E,L+E], f^{(\mathtt{Lap})}_{x,\sigma}(y) = \frac{1}{\lambda^{(\mathtt{Lap})}_{L,E,\sigma}} e^{-\frac{\min(|y-x|,L)}{\sigma}}$ where $\lambda^{(\mathtt{Lap})}_{L,E,\sigma}$ is chosen such that $\int_{-E-L}^{E+L} f^{(\mathtt{Lap})}_{x,\sigma}(y) dy = 1.$
\end{itemize}
The following proposition provides the value of the coefficient $\lambda^{(\mathtt{Lap})}_{L,E,\sigma}.$
\begin{prop}\label{prop:TruncLapCoef}
    For any positive values $L,E,\sigma$ and $x\in [-E,E],$ we have $\lambda^{(\mathtt{Lap})}_{L,E,\sigma}=2\left(\sigma+(E-\sigma)e^{-\frac{L}{\sigma}}\right).$
\end{prop}
\begin{proof}
    Proof is provided in Appendices.
\end{proof}
\subsection{Privacy}
The truncated Laplace mechanism satisfies $\epsilon$-differential privacy, and the details are shown in the following Proposition.
\begin{prop}\label{prop:TruncLapPriv}
Let $L,E,$ and $\sigma$ be an arbitrary triple of positive values and consider the mechanism $\mathcal{M}^{(\mathtt{Lap})}_{L,E,\sigma}:[-E,E]\rightarrow [-E-L,E+L]$ such that for any $x\in [-E,E], \mathcal{M}_{L,E,\sigma}^{(\mathtt{Lap})}(x)=y$ where $y$ is sampled following the distribution $f^{(\mathtt{Lap})}_{x,\sigma}.$ Then for any privacy budget $\epsilon>0,$ such mechanism satisfies $\epsilon$-Differential Privacy if $\sigma\geq \frac{L}{\epsilon}.$
\end{prop}
\begin{proof}
    Note that for any $x\in [-E,E],$ the distribution $f^{(\mathtt{Lap})}_{x,\sigma}$ achieves its maximum of $\frac{1}{\lambda^{(\mathtt{Lap})}_{L,E,\sigma}}$ when $y=x$ and its minimum of $\frac{1}{\lambda^{(\mathtt{Lap})}_{L,E,\sigma}} e^{-\frac{L}{\sigma}}$ whenever $|y-x|\geq L.$ Hence for any $y\in [-E-L,E+L]$ and $x_1,x_2\in [-E,E],$ letting $\sigma\geq\frac{L}{\epsilon},$ we have
    \begin{eqnarray*}
        \frac{f^{(\mathtt{Lap})}_{x_1,\sigma}(y)}{f^{(\mathtt{Lap})}_{x_2,\sigma}(y)}&=&\frac{e^{-\frac{\min(|y-x_1|,L)}{\sigma}}}{e^{-\frac{\min(|y-x_2|,L)}{\sigma}}}\\
        &\leq & \frac{1}{e^{-\frac{L}{\sigma}}}=e^{\frac{L}{\sigma}}\leq e^{\epsilon},
    \end{eqnarray*}
    which proves the privacy claim.
\end{proof}
\subsection{Accuracy}
Here for positive values $L,E,$ and privacy budget $\epsilon>0,$ we denote by $\mathcal{M}^{(\mathtt{Lap},\epsilon)}_{L,E} = \mathcal{M}^{(\mathtt{Lap})}_{L,E,\frac{L}{\epsilon}}.$ For simplicity of notation, we denote it by $\mathcal{M}$ where $L,E,$ and $\epsilon$ are assumed to be fixed. Next we use $y$ as an estimator of $x$ given the perturbation mechanism $\mathcal{M}.$
\begin{prop}\label{prop:TruncLapEst}
    Let $E$ be a positive value such that for any data value $x,$ we have $x\in [-E,E].$ We further let $\epsilon>0$ be a privacy budget. Then there exists $k^\ast \in(1,2)$ such that  by letting $\sigma = \frac{E}{k^\ast}$ and $L=\sigma \epsilon,$ the following holds. Given $y=\mathcal{M}(x), y$ is a biased estimator of $x$ with expected value 
    \[
    \mathbb{E}(y)=x\left(\frac{1-(1+\epsilon)e^{-\epsilon}}{1-\left(1-\frac{E}{\sigma}\right)e^{-\epsilon}}\right)
    \]
    and expected squared distance 
    \[
            \xi\triangleq \mathbb{E}((y-x)^2)= \frac{2\sigma^2 + x^2 e^{-\epsilon}\left(\epsilon+\frac{E}{\sigma}\right)}{1-\left(1-\frac{E}{\sigma}\right)e^{-\epsilon}}. 
    \]
\end{prop}
\begin{proof}
    A full proof is provided in the supplementary material due to page limits.
\end{proof}

\section{Truncated Discrete Laplace Mechanism: A Laplace-inspired Perturbation}\label{sec:CircPSISol-TruncDisLapGaus}
This section focuses on our first Laplace-inspired DP mechanism, namely truncated discrete Laplace (TDL), by further modifying the truncated Laplace distribution to have a discrete support. More specifically, instead of having the sample space to be $[-L-E,L+E],$ we assume that the sample space is $[-L-E,L+E]\cap 2^{-p}\mathbb{Z}$ where $p$ is the precision parameter and $2^{-p}\mathbb{Z}=\left\{\frac{a}{2^p}: a\in \mathbb{Z}\right\}.$ For simplicity of notation, for any bound $B\in \mathbb{R}_{>0}$ and precision parameter $p\in \mathbb{Z}_{>0},$ we denote by $\mathcal{A}_{p,B}\triangleq [-B,B]\cap 2^{-p}\mathbb{Z}.$ Here we also assume that for any value $x,E,L,$ there are no rounding error during the computation, that is, $x,E,L\in 2^{-p}\mathbb{Z}.$

\subsection{Definition of the Mechanism}\label{sec:CircPsiSol-TruncDisLapGaus-TruncDisLap}
Given the truncated Laplace distribution $f^{(\mathrm{Lap})}_{x,\sigma}$ as defined in the preceding section, we define the truncated discrete Laplace mechanism as follows.
\begin{defn}[Truncated Discrete Laplace Mechanism $\mathcal{M}^{(\mathtt{DLap})}_{L,E,\sigma}$]
    Given a data $x\in \mathcal{A}_{p,E},$ it is perturbed to $y\in \mathcal{A}_{p,L+E}$ with probability $f^{(\mathtt{DLap})}_{x,\sigma}(y)=\frac{1}{\lambda^{(\mathtt{DLap})}_{L,E,\sigma,p}}e^{-\frac{-\min(|y-x|,L)}{\sigma}}$ where $\lambda^{(\mathtt{DLap})}_{L,E,\sigma,p}$ is chosen such that $\sum_{y\in \mathcal{A}_{p,L+E}} f^{(\mathtt{DLap})}_{x,\sigma}(y) =1.$
\end{defn}
 It is easy to see that in this mechanism the noise is sampled in the truncated and discrete support $\mathcal{A}_{p,B}$. For any $x\in \mathcal{A}_{p,E},$ we partition $\mathcal{A}_{p,L+E}$ to $5$ sets depending on its position with respect to $x,$ namely
    \begin{enumerate}
        \item $T_0^{(x)}=\{x\}.$ Note that $f^{(\mathtt{DLap})}_{x,\sigma}(x)=\frac{1}{\lambda_{L,E,\sigma,p}^{(\mathtt{DLap})}}$
        \item $T_1^{(x)}=[-L-E,x-L)\cap 2^{-p}\mathbb{Z}=\{x-L-i2^{-p}:i=1,\cdots, 2^p(x+E)\}.$ Note that for $y\in T_1^{(x)}, f^{(\mathtt{DLap})}_{x,\sigma}(x)=\frac{1}{\lambda_{L,E,\sigma,p}^{(\mathtt{DLap})}}e^{-\frac{L}{\sigma}}.$
        \item $T_2^{(x)}=[x-L,x)\cap 2^{-p}\mathbb{Z}=\{x-i2^{-p}:i=1,\cdots, 2^pL\}.$ Note that for $y\in T_2^{(x)}, f^{(\mathtt{DLap})}_{x,\sigma}(y)=\frac{1}{\lambda_{L,E,\sigma,p}^{(\mathtt{DLap})}}e^{-\frac{x-y}{\sigma}}.$
        \item $T_3^{(x)}=(x,x+L]\cap 2^{-p}\mathbb{Z}=\{x+i2^{-p}:i=1,\cdots, 2^pL\}.$ Note that for $y\in T_3^{(x)}, f^{(\mathtt{DLap})}_{x,\sigma}(y)=\frac{1}{\lambda_{L,E,\sigma,p}^{(\mathtt{DLap})}}e^{-\frac{y-x}{\sigma}}.$
        \item $T_4^{(x)}=(x+L,L+E]\cap 2^{-p}\mathbb{Z} = \{x+L+i2^{-p}:i=1,\cdots, 2^p(E-x)\}.$ Note that for $y\in T_4^{(x)}, f^{(\mathtt{DLap})}_{x,\sigma}(y)=\frac{1}{\lambda_{L,E,\sigma,p}^{(\mathtt{DLap})}}e^{-\frac{L}{\sigma}}.$
    \end{enumerate}
 We next give the value of $\lambda_{L,E,\sigma,p}^{(\mathtt{DLap})}.$
    \begin{prop}\label{prop:TruncDisLapCoef}
        For any positive values $L,E,\sigma, p$ and $x\in \mathcal{A}_{p,E},$ we have 
        \[
        \lambda^{(\mathtt{DLap})}_{L,E,\sigma,p}=2^{p+1}Ee^{-\frac{L}{\sigma}} + 1 + 2 \frac{1-e^{-\frac{L}{\sigma}}}{e^{\frac{2^{-p}}{\sigma}}-1}.
        \]
    \end{prop}
    \begin{proof}
    It is easy to see that $\lambda \triangleq \lambda^{(\mathtt{DLap})}_{L,E,\sigma,p} = \sum_{y\in \mathcal{A}_{p,L+E}} e^{-\frac{\min(|x-y|,L)}{\sigma}}.$ For $j=0,\cdots, 4,$ let $\lambda_j=\sum_{y\in T_j^{(x)}} e^{-\frac{\min(|x-y|,L)}{\sigma}}.$
    It is then easy to see that $\lambda_0 = 1, \lambda_1 = 2^p(x+E)e^{-\frac{L}{\sigma}},$ and $\lambda_4= 2^p(E-x) e^{-\frac{L}{\sigma}}.$
    On the other hand, we have 
    \[\lambda_2 = \sum_{i=1}^{2^pL} e^{-\frac{i2^{-p}}{\sigma}}\]
    and
    \[\lambda_3 = \sum_{i=1}^{2^pL} e^{-\frac{i2^{-p}}{\sigma}}.\]
    Thus, $\lambda$ is the sum of all $\lambda_i$ for $i=0,\cdots,4$, concluding the proof. 
\end{proof}

\subsection{Privacy}
Regarding privacy, the proposed truncated discrete Laplace mechanism achieves $(\epsilon,0)$-DP, whereas most existing work of this kind mostly cannot avoid a probability of failure in their privacy guarantees. We give the privacy argument in the following proposition without proof which is essentially the same as that of Proposition~\ref{prop:TruncLapPriv}.

\begin{prop}\label{prop:TruncDisLapPriv}
        Let $L,E,\sigma,$ and $p$ be an arbitary $4$-tuple of positive values and consider the mechanism $\mathcal{M}^{(\mathtt{DLap})}_{L,E,\sigma,p}: \mathcal{A}_{p,E}\rightarrow \mathcal{A}_{p,L+E}$ such that for any $x\in \mathcal{A}_{p,E},\mathcal{M}^{(\mathtt{DLap})}_{L,E,\sigma,p}(x) = y$ with probability mass function $f_{x,\sigma}^{(\mathtt{DLap})}(y).$ Then for any privacy budget $\epsilon>0,$ such mechanism satisfies $\epsilon-$Differential Privacy if $\sigma\geq \frac{L}{\epsilon}.$
\end{prop}
Note that up to now, we only consider the strict differential privacy when considering the privacy guarantee of our perturbation mechanism, which provides uniform protection of any value against any other value in the input space. However, depending on the application, such uniform protection may not be necessary. Instead, at some scenario, such as location information, privacy concern is higher in protecting among other values with closer proximity compared to those with further distance from our value. In such case, the noise necessary to provide such protection may become smaller. This notion of distance-based protection was first proposed in \cite{ABCP13}, called geo-indistinguishability. In the following we provide a short review of the concept of distance-based privacy, denoted by $d_\chi$-privacy. Then, we demonstrate that our proposed truncated discrete Laplace mechanism satisfies $d_\chi$-privacy in Proposition \ref{prop:TruncDisLapdchiPriv}.

\begin{defn}\label{def:dchiDPdef}
    Let $\chi$ be a metric space containing all possible values of the user's data and $\mathcal{R}$ be a report space containing all possible reports the user may send to the server. Let $\mathcal{M}:\chi\rightarrow\mathcal{R}$ be a randomized mechanism a user may use to generate a report based on his data. Then $\mathcal{M}(\mathbf{x})$ is a random variable following the distribution defined by the probability density function we denote by $f_\mathbf{x}:\mathcal{R}\rightarrow \mathbb{R}_{\geq 0}.$ The mechanism $\mathcal{M}$ is said to be $d_\chi$-private if and only if for any $\mathbf{x},\mathbf{x}'\in \chi$ and $\mathbf{y}\in \mathcal{R},$ 
    \[
    f_\mathbf{x}(\mathbf{y})\leq e^{d_\chi(\mathbf{x},\mathbf{x}')}f_{\mathbf{x}'}(\mathbf{y}).
    \]
    Intuitively, the statistical distance between two distributions outputted by a $d_\chi$-private mechanism with respect to two distinct data decreases as the distance between the two data decreases with respect to the distance function $d_\chi.$ In order to incorporate a measure of privacy loss, we may embed the privacy budget to the distance function. More specifically, instead of having $d_\chi$ to be the original distance function, we have $d_\chi(\mathbf{x},\mathbf{x}')=\epsilon d(\mathbf{x},\mathbf{x}')$ where $d(\cdot,\cdot)$ is the original distance function. Therefore, for a fixed $\epsilon>0,$ a mechanism $\mathcal{M}$ is said to have $d_\chi$-privacy if for any $\mathbf{x},\mathbf{x}'\in \chi$ and $\mathbf{y}\in \mathcal{R},$
    \[ f_\mathbf{x}(\mathbf{y})\leq e^{\epsilon d(\mathbf{x},\mathbf{x}')}f_{\mathbf{x}'}(\mathbf{y}).
    \]
    \end{defn}
In this work, we consider $\chi=\mathcal{A}_{p,E}$ and $\mathcal{R}=\mathcal{A}_{p,L+E}$ equipped with the standard $\ell_1$ distance.

Based on this definition, we have the following proposition
\begin{prop}\label{prop:TruncDisLapdchiPriv}
        Let $L,E,\sigma,$ and $p$ be an arbitary $4$-tuple of positive values and consider the mechanism $\mathcal{M}^{(\mathtt{DLap})}_{L,E,\sigma,p}: \mathcal{A}_{p,E}\rightarrow \mathcal{A}_{p,L+E}$ such that for any $x\in \mathcal{A}_{p,E},\mathcal{M}^{(\mathtt{DLap})}_{L,E,\sigma,p}(x) = y$ with probability mass function $f_{x,\sigma}^{(\mathtt{DLap})}(y).$ Then for any privacy budget $\epsilon>0,$ such mechanism satisfies $d_\chi$-privacy if $\sigma\geq \frac{1}{\epsilon}.$
\end{prop}
\begin{proof}
        Let $x,y\in \mathcal{A}_{p,E}.$ Then for $z\in \mathcal{A}_{p,L+E},$ we have $\frac{f_{x,\sigma}^{(\mathtt{DLap})}(z)}{f_{y,\sigma}^{(\mathtt{DLap})}(z)}=\frac{e^{-\frac{\min(|z-x|,L)}{\sigma}}}{e^{-\frac{\min(|z-y|,L)}{\sigma}}}.$ We will discuss the ratio in three cases depending on the distances between $x,y$ and $z.$
        \begin{enumerate}
            \item If $|z-x|\geq L$ and $|z-y|\geq L,$ then the ratio equals $1$ which is upper bounded by $e^{\epsilon |x-y|}$ for any $\epsilon> 0.$
            \item If $|z-y|\leq L\leq |z-x|,$ then the ratio becomes $e^{\frac{|z-y|-L}{\sigma}}\leq 1$ which is again upper bounded by $e^{\epsilon |x-y|}$ for any $\epsilon>0.$
            \item If $|z-x|\leq L,$ then the ratio equals $e^{\frac{\min(|z-y|,L)-|z-x|}{\sigma}}\leq e^{\frac{|z-y|-|z-x|}{\sigma}}\leq e^{\frac{|y-x|}{\sigma}}.$ Hence the ratio is at most $e^{\epsilon |x-y|}$ as long as $\sigma\geq \frac{1}{\epsilon}.$
        \end{enumerate}
        This shows that in any case, if we set $\sigma \geq\frac{1}{\epsilon},$ the ratio is at most $e^{\epsilon |x-y|},$ proving that the mechanism satisfies $d_\chi$-privacy.
\end{proof}
\subsection{Accuracy}
    Given the truncated discrete mechanism, we use $y=\mathcal{M}^{(\mathtt{DLap})}_{L,E,\sigma,p}(x)$ to estimate $x$ and analyze its variance. Here we simplify the notation of the mechanism to $\mathcal{M}.$
    \begin{prop}\label{prop:TruncDisLapEst}
        Let $E$ and $p$ be positive values such that for any data value $x,$ we have $x\in \mathcal{A}_{p,E}.$ We further let $\epsilon>0$ be a privacy budget. Then assuming that $L=O(E),$ regardless of the value of $\sigma,$ given $y=\mathcal{M}(x),$ $y$ is a biased estimator of $x$ with expected value 
        \begin{align*}
        &\mathbb{E}(y) = \\
        &x\left(\frac{(1-e^{-\frac{L}{\sigma}})(1+e^{\frac{2^{-p}}{\sigma}}) - 2e^{-\frac{L}{\sigma}} 2^p L (e^{\frac{2^{-p}}{\sigma}}-1)}
        {(1-e^{-\frac{L}{\sigma}})(e^{\frac{2^{-p}}{\sigma}}+1)+ e^{-\frac{L}{\sigma}}(e^{\frac{2^{-p}}{\sigma}}-1)(1+2\cdot 2^pE)}\right),
        \end{align*}

        and expected squared error
         \begin{align*}
            & \xi \triangleq \mathbb{E}((y-x)^2) \\
           &= x^2\cdot\frac{2+e^{-\frac{L}{\sigma}}(1+2\cdot 2^p(E+L))}{\left(1-e^{-\frac{L}{\sigma}}\right)\frac{1+e^{\frac{2^{-p}}{\sigma}}}{e^{\frac{2^{-p}}{\sigma}}-1}+e^{-\frac{L}{\sigma}}(1+2\cdot 2^p E)}\\
               &+e^{-\frac{L}{\sigma}}\frac{2\cdot 2^p\left(L^2E+LE^2+\frac{E^3}{3}\right) + \left(E^2+2LE-\frac{2L^2}{e^{\frac{2^{-p}}{\sigma}}-1}\right)}{\left(1-e^{-\frac{L}{\sigma}}\right)\frac{e^{\frac{2^{-p}}{\sigma}}+1}{e^{\frac{2^{-p}}{\sigma}}-1} + e^{-\frac{L}{\sigma}} (1+2\cdot 2^pE)}\\
            &+\frac{e^{-\frac{L}{\sigma}}2^{-p}\left(\frac{E}{3} - \frac{4Le^{\frac{2^{-p}}{\sigma}}}{(e^{\frac{2^{-p}}{\sigma}}-1)^2}\right)}{\left(1-e^{-\frac{L}{\sigma}}\right)\frac{e^{\frac{2^{-p}}{\sigma}}+1}{e^{\frac{2^{-p}}{\sigma}}-1}+ e^{-\frac{L}{\sigma}} (1+2\cdot 2^pE)} \\
            &+ \frac{2\cdot 2^{-2p}(1-e^{-\frac{L}{\sigma}})\frac{e^{\frac{2\cdot 2^{-p}}{\sigma}}+e^{\frac{2^{-p}}{\sigma}}}{(e^{\frac{2^{-p}}{\sigma}}-1)^3}}{\left(1-e^{-\frac{L}{\sigma}}\right)\frac{e^{\frac{2^{-p}}{\sigma}}+1}{e^{\frac{2^{-p}}{\sigma}}-1} + e^{-\frac{L}{\sigma}} (1+2\cdot 2^pE)} \\
        \end{align*}
    \end{prop}
    \begin{proof}
        A full proof is provided in the supplementary material due to page limits.
    \end{proof}
\begin{rmk}\label{rem:DLapUtil}
        Note that the calculation will be done over a finite field $\mathbb{F}_q$ of size $q\geq |\mathcal{A}_{p,L+E}|=2\cdot 2^p (L+E).$ Hence assuming that $L=O(E),$ Proposition \ref{prop:TruncDisLapEst} asserts that the estimator $y,$ its expected bias is 
        \begin{align*}
         & |\mathbb{E}(y)-x| = \\
         & x\left(\frac{e^{-\frac{L}{\sigma}}\left(e^{\frac{2^{-p}}{\sigma}}-1\right)(1+2\cdot 2^p(E+L))}{(1-e^{-\frac{L}{\sigma}})(e^{\frac{2^{-p}}{\sigma}}+1)+ e^{-\frac{L}{\sigma}}(e^{\frac{2^{-p}}{\sigma}}-1)(1+2\cdot 2^pE)}\right),    
        \end{align*}
        with expected squared error $O\left(2^{-2p}q^2\right).$ Furthermore, by setting $\sigma= \frac{L}{\epsilon}$ and $L=O(E),$ the expected squared error of the estimator is $O(\sigma^2).$
        \end{rmk}
\section{Truncated Cumulative Laplace Mechanism: A variant of TDL}
Along with the truncated discrete Laplace noise we defined in the preceding section, we give our second Laplace-inspired perturbation, referred to as the truncated cumulative Laplace (TCL) distribution. Intuitively, instead of simply taking the density function value of the original Laplace distribution in each sample, we set the sample to be the report if the original sample is in the interval of length $2^{-p}$ immediately after the discrete value. Such variant is discussed in detail in this section.

\subsection{Definition of the Mechanism}
In this section, we define a DP mechanism defined with the TCL distribution. 
\begin{defn}[Truncated Cumulative Laplace Mechanism $\mathcal{M}^{(\mathtt{CLap})}_{L,E,\sigma,p}$]
     Given a data $x\in \mathcal{A}_{p,E},$ it is perturbed to $y\in \mathcal{B}_{p,L+E}\triangleq \mathcal{A}_{p,L+E}\setminus\{L+E\}$ with probability 
    \[f^{(\mathtt{CLap})}_{x,\sigma}(y)=
    \frac{1}{\lambda^{(\mathtt{CLap})}_{L,E,\sigma}}\int_{y}^{y+2^{-p}}e^{-\frac{\min(|r-x|,L)}{\sigma}}dr,\]
    
where $\lambda^{(\mathtt{CLap})}_{L,E,\sigma}$ is chosen such that $\sum_{y\in \mathcal{B}_{p,L+E}} f^{(\mathtt{CLap})}_{x,\sigma}(y) =1.$
\end{defn}

 
 Note that the sampled value equals $L+E$ with probability $0.$ Because of this, the sample space becomes $\mathcal{B}_{p,L+E}$ where $L+E$ is impossible to be sampled. 
 \begin{prop}\label{prop:TrunCLapCoef}
        For any positive values $L,E,\sigma, p$ and $x\in \mathcal{A}_{p,E},$ we have 
        \[
    \lambda_{L,E,\sigma}^{(\mathtt{CLap})}=2\left(\sigma(1-e^{-\frac{L}{\sigma}})+e^{-\frac{L}{\sigma}}E\right).
        \]
    \end{prop}
 \begin{proof}[Sketch of Proof] As per the definition of $f^{(\mathtt{CLap})}_{(x,\sigma)}$, it is derived from a truncated Laplace distribution $f^{(\mathtt{Lap})}_{(x,\sigma)}$ in a way that the integral of $f^{(\mathtt{Lap})}_{(x,\sigma)}$ over the interval $[y,y+2^{-p}]$ is equal to the probability of occurrence of of sampling $y$. Therefore, we have
 \[
    \lambda_{L,E,\sigma}^{(\mathtt{CLap})}=\lambda_{L,E,\sigma}^{(\mathtt{Lap})}.
 \]
     
 \end{proof}
 
\subsection{Privacy}
The truncated cumulative Laplace mechanism $\mathcal{M}^{(\mathtt{CLap})}_{L,E,\sigma,p}$ achieves $\epsilon$-DP.
\begin{prop}\label{prop:TruncDisCLapPriv}
    Let $L,E,\sigma,$ and $p$ be an arbitary $4$-tuple of positive values and consider the mechanism $\mathcal{M}^{(\mathtt{CLap})}_{L,E,\sigma,p}: \mathcal{A}_{p,E}\rightarrow \mathcal{B}_{p,L+E}$ such that for any $x\in \mathcal{A}_{p,E},\mathcal{M}^{(\mathtt{CLap})}_{L,E,\sigma,p}(x) = y$ with probability mass function $f_{x,\sigma}^{(\mathtt{CLap})}(y).$ Then for any privacy budget $\epsilon>0,$ such mechanism satisfies $\epsilon-$Differential Privacy if $\sigma\geq \frac{L}{\epsilon}.$
\end{prop}
 \begin{proof}
    Note that for any $x\in \mathcal{A}_{p,E},$ the distribution $f^{(\mathtt{CLap})}_{x,\sigma}$ achieves its maximum $M$ when $y=x,$ which is
    \[M= \frac{1}{\lambda_{L,E,\sigma}^{(\mathtt{CLap})}} \int_{x}^{x+2^{-p}} e^{-\frac{r-x}{\sigma}}dr=\frac{\sigma}{\lambda_{L,E,\sigma}^{(\mathtt{CLap})}}\left(1-e^{-\frac{2^{-p}}{\sigma}}\right).\]
    Note that we have $e^{-\frac{2^{-p}}{\sigma}}=\sum_{j\geq 0} \frac{1}{j!}\cdot \left(-\frac{2^{-p}}{\sigma}\right)^j\geq 1-\frac{2^{-p}}{\sigma}.$ Hence we have $M\leq \frac{2^{-p}}{\lambda_{L,E,\sigma}^{(\mathtt{CLap})}}.$ 

    On the other hand, $f_{x,\sigma}^{(\mathtt{CLap})}$ achieves its minimum $m$ when $y=y^\ast$ such that for any $r\in [y,y+2^{-p}], |r-x|>L,$ that is,
    \[m=\frac{1}{\lambda_{L,E,\sigma}^{(\mathtt{CLap})}} \int_{x}^{x+2^{-p}} e^{-\frac{L}{\sigma}}dr=\frac{2^{-p}e^{-\frac{L}{\sigma}}}{\lambda_{L,E,\sigma}^{(\mathtt{CLap})}}.\]
    Hence for any $y\in [-E-L,E+L]$ and $x_1,x_2\in [-E,E],$ letting $\sigma\geq\frac{L}{\epsilon},$ we have
    \begin{eqnarray*}
        \frac{f^{(\mathtt{Lap})}_{x_1,\sigma}(y)}{f^{(\mathtt{Lap})}_{x,\sigma}(y)}&\leq&\frac{M}{m}=e^{\frac{L}{\sigma}}\leq e^{\epsilon}
    \end{eqnarray*}
    which proves the privacy claim.
\end{proof}
Similar as the previous section, we consider the mechanism under $d_{\chi}$-privacy definition.
\begin{prop}\label{prop:TruncDisCLapdchiPriv}
        Let $L,E,\sigma,$ and $p$ be an arbitrary $4$-tuple of positive values and consider the mechanism $\mathcal{M}^{(\mathtt{CLap})}_{L,E,\sigma,p}: \mathcal{A}_{p,E}\rightarrow \mathcal{B}_{p,L+E}$ such that for any $x\in \mathcal{A}_{p,E},\mathcal{M}^{(\mathtt{CLap})}_{L,E,\sigma,p}(x) = y$ with probability mass function $f_{x,\sigma}^{(\mathtt{CLap})}(y).$ Then for any privacy budget $\epsilon>0,$ such mechanism satisfies $d_{\chi}$- privacy if $\sigma\geq \max\left(\frac{2}{\epsilon},2^{-p}\right).$
    \end{prop}
    \begin{proof}
        A full proof is provided in the supplementary material due to page limits.
    \end{proof}
\subsection{Accuracy}
Lastly we use $y=\mathcal{M}_{L,E,\sigma,p}^{(\mathtt{CLap})}(x)$ as an estimator for $x$ and we analyze its variance. For simplicity of notation, we denote the mechanism by $\mathcal{M}.$

\begin{prop}\label{prop:TruncDisCLapEst}
        Let $E$ and $p$ be positive values such that for any data value $x,$ we have $x\in \mathcal{A}_{p,E}.$ We further let $\epsilon>0$ be a privacy budget. Then assuming that $L=O(E)$ and $\sigma = O(E^2),$ given that $y=\mathcal{M}(x),$ it is a biased estimator of $x$ with expected value 
        \[\mathbb{E}\triangleq \mathbb{E}(y)=x\left(\frac{\sigma(1-e^{-\frac{L}{\sigma}})-e^{-\frac{L}{\sigma}}L}{\sigma(1-e^{-\frac{L}{\sigma}})+e^{-\frac{L}{\sigma}}E}\right)-\frac{2^{-p}}{2}\]
        and expected squared error $\xi\triangleq \mathbb{E}((y-x)^2)=O(E^2).$
    \end{prop}
    \begin{proof}
        For the full proof, please refer to the supplementary material.
    \end{proof}
\begin{rmk}
We note that Remark \ref{rem:DLapUtil} also applies to the proposed mechanism in this section.    
\end{rmk}
\section{Multi-Party Sampling of the Perturbation}\label{sec:sampling}
In this section, we discuss how we may perturb the data
$x\in [-E,E]$ with the noise following the two proposed distributions in the scenario where $x$ is secretly shared and the noise needs to be injected before the value of $x$ is revealed. For ease of notation, we assume that $x$ and $\sigma$ are fixed and the distributions of the perturbed data, $f_{x,\sigma}^{(\mathtt{DLap})}$ and $f_{x,\sigma}^{(\mathtt{CLap})},$ are denoted by $f^{(D)}$ and $f^{(C)}$ respectively. We also denote the constant multiplier in $f^{(D)}$ and $f^{(C)}$ by $\lambda^{(D)}$ and $\lambda^{(C)}$, respectively. 
\subsection{Parameters}
First, we will express $f^{(C)}$ and $f^{(D)}$ in a different way to separate the influence of $x$ to the noise sampling. Once that is achieved, we propose the sampling method for the noise in 2PC. We define the following notation.
\begin{defn}\label{def:SepTool}
    Let $m,p,B$ be positive integers, $x\in 2^{-p}\mathbb{Z}$ and $\sigma\in \mathbb{R}_{>0}$ We define the following distributions:
    \begin{itemize}
        \item For a set $S\subseteq \mathbb{R}$ of size $m,$ define $g^{(U,S)}:S\rightarrow \mathbb{R}$ such that $g^{(U,S)}(y)=\frac{1}{m}.$ That is, $g^{(U,S)}$ represents the uniform distribution with sampling space $S.$
        \item Let $S=[x-B,x+B]\cap 2^{-p}\mathbb{Z}.$ We define $g^{(D,S)}:S\rightarrow \mathbb{R}$ such that $g^{(D,S)}(y)= \frac{1}{\lambda^{(D,S)}} e^{-\frac{|y-x|}{\sigma}}$ where $\lambda^{(D,S)}\triangleq \sum_{y\in S} e^{-\frac{|y-x|}{\sigma}}.$ In here, $g^{(D,S)}$ represents the standard discrete Laplace distribution with mean $x,$ parameter $\sigma,$ and sampling space $S.$ We further denote by $g^{(D,B)}(y)$ to be the distribution when $x=0.$ It is easy to see that to sample from $g^{(D,S)},$ we can sample from $g^{(D,B)}$ and add $x$ to its output. This observation allows the sampling of $g^{(D,S)}$ to first be done independently of $x$ before being added to $x$ in a latter stage.

        Similarly, for $T=[x-B,x+B)\cap 2^{-p}\mathbb{Z},$ we define $g^{(C,T)}:T\rightarrow \mathbb{R}$ as $g^{(C,T)}(y) = \frac{1}{\lambda^{(C,T)}} \int_y^{y+2^{-p}} e^{-\frac{|r-x|}{\sigma}}dr$ and $g^{(C,B)}$ as the distribution when $x=0.$ We again have the same observation that sampling from $g^{(C,T)}$ can be done by first sampling from $g^{(C,B)}$ and add the output by $x.$
    \end{itemize}
\end{defn}

In our discussion, we assume the existence of a secure 2PC protocol $\Pi^{(\mathtt{BerSample})}(p)$ which, given the secret share of a real number $p\in [0,1],$ outputs the secret share of the fixed point encoding of $1$ with probability $p$ and the fixed point encoding of $0$ with probability $1-p.$ Similarly, we also assume the existence of a secure 2PC protocol $\Pi^{(\mathtt{Uni})}(M)$ which outputs the secret share of an element $z\in\{0,\cdots, M-1\}$ with uniform probability.

\subsection{2-PC Perturbation of $f^{(D)}$}
We note that $f^{(D)}$ can be seen as a combination of two distributions; the uniform distribution with $2^p(2E)$ possible values ($y\in S\triangleq ([-L-E,x-L)\cup (x+L,L+E])\cap 2^{-p}\mathbb{Z}$) which follows $g^{(U,S)}$ and $g^{(D,S)}$ when $y\in S\triangleq [x-L,x+L]\cap 2^{-p}\mathbb{Z}.$ We propose two functionalities we denote by $\mathcal{F}^{(O)}(M)$ and $\mathcal{F}^{(I,D)}(p,L).$ The first functionality $\mathcal{F}^{(O)}(M)$ takes a positive integer $M$ as an input and it outputs a random element of $\{0,\cdots, M-1\}$ where each element is sampled with uniform probability $\frac{1}{M}.$ On the other hand, the second functionality $\mathcal{F}^{(I,D)}(p,L)$ takes the bound and precision parameters $L$ and $p,$ while outputting a random element of $[-L,L]\cap 2^{-p}\mathbb{Z}$ where for any element $y,$ it is sampled with probability $g^{(D,L)}(y).$ 

To realize the perturbation mechanism, we propose a two-step sampling for $f^{(D)}$. At the first step, a functionality $\mathcal{F}^{(D)}.\mathtt{Noise}$ produces the noise independent of $x$, while at the second step the other functionality $\mathcal{F}^{(D)}.\mathtt{Perturb}$ takes the output of $\mathcal{F}^{(D)}.\mathtt{Noise}$ and the value of $x$ to generate an output following the distribution $f^{(D)}.$ 

More specifically, the two phases work as follows. In the first phase,  $\mathcal{F}^{(D)}.\mathtt{Noise}$ yields the noise independent of input value $x$. Note that regardless of the value of $x,$ the number of possible outputs $y$ such that $|y-x|>L$ is always the same, which is $2\cdot 2^pE.$ Furthermore, the probability of choosing such values is also always the same, which is $2\cdot 2^p E \frac{e^{-\frac{L}{\sigma}}}{\lambda^{(D)}}.$ Because of this observation, with a probability of $2\cdot 2^p E \frac{e^{-\frac{L}{\sigma}}}{\lambda^{(D)}},$  $\mathcal{F}^{(D)}.\mathtt{Noise}$ invokes $\mathcal{F}^{(O)}(M)$ which samples from the uniform distribution $g^{(U,S)}$ over the support $S\triangleq \{0,\cdots, 2\cdot 2^pE-1\}$. Otherwise, $\mathcal{F}^{(D)}.\mathtt{Noise}$ invokes $\mathcal{F}^{(I,D)}(p,L)$ which samples from $g^{(D,L)}$ over the support $[-L,L]\cap 2^{-p}\mathbb{Z}$. 
Once $\mathcal{F}^{(D)}.\mathtt{Noise}$ is done, we are ready to perturb the value of $x$. If the noise comes from the uniform distribution, we use the natural bijection $\phi:\{0,\cdots, 2\cdot 2^pE-1\}\rightarrow ([-L-E,x-L)\cup (x+L,L+E])\cap 2^{-p}\mathbb{Z}$ to transform the noise to the appropriate value. On the other hand, if the noise comes from the $g^{(D,L)}$, we simply add $x$ to the noise to obtain a perturbed $x$ following $g^{(D,[x-L,x+L]\cap 2^{-p}\mathbb{Z})}.$ It is then easy to verify that the resulting value follows $f^{(D)}.$

We describe the procedures of $\mathcal{F}^{(D)}.\mathtt{Noise}$ in Algorithm \ref{func:FDNoise}.
\begin{algorithm}[htbp]
\caption{$(i,y) \leftarrow \mathcal{F}^{(D)}.\mathtt{Noise}()$}
\label{func:FDNoise}
\begin{algorithmic}[1]
\REQUIRE Let $E,L,p,\sigma$ be as previously defined. 
\ENSURE The functionality outputs $(i,y)$ where $i=0$ if $y$ is the output of $\mathcal{F}^{(O)}(2\cdot 2^pE)$ and $i=1$ if $y$ is the output of $\mathcal{F}^{(I,D)}(p,L).$ 
\STATE Sample $i\in \{0,1\}$ following a Bernoulli distribution with success probability $1-\frac{2\cdot 2^pE e^{-\frac{L}{\sigma}}}{\lambda^{(D)}};$
\IF{$i=0$}
\STATE Sample $a\in\{0,\cdots, 2\cdot 2^pE-1\}$ uniformly at random;
\STATE Return $(i,a);$
\ELSE
\STATE Sample $b\in [-L,L]\cap 2^{-p}\mathbb{Z}$ following the distribution $g^{(D,L)};$
\STATE Return $(i,b);$
\ENDIF
\end{algorithmic}
\end{algorithm}

Furthermore, the specification of $\mathcal{F}^{(D)}.\mathtt{Perturb}$ can be found in Algorithm \ref{func:FDPerturb}.
\begin{algorithm}[htbp]
\caption{$z \leftarrow \mathcal{F}^{(D)}.\mathtt{Perturb}(x,i,y)$}
\label{func:FDPerturb}
\begin{algorithmic}[1]
\REQUIRE Let $x\in [-E,E]\cap 2^{-p}\mathbb{Z}$ be the value we want to perturb and $(i,y)$ is the output of $\mathcal{F}^{(D)}.\mathtt{Noise}()$ 
\ENSURE The functionality outputs $z$ following the distribution $f^{(D)}$
\IF{$i==0$}
\IF{$y\leq 2^p(E+x)-1$} 
\STATE Define $z\leftarrow -L-E + y2^{-p};$
\ELSE
\STATE Define $z\leftarrow L-E + 2^{-p}(y+1);$
\ENDIF
\ELSE
\STATE Define $z\leftarrow x+y;$
\ENDIF
\STATE Return $z;$
\end{algorithmic}
\end{algorithm}
We then propose the secure protocols realizing $\mathcal{F}^{(D)}.\mathtt{Noise}$ and $\mathcal{F}^{(D)}.\mathtt{Perturb},$ denoted by $\Pi^{(D)}.\mathtt{Noise}$ and $\Pi^{(D)}.\mathtt{Perturb}$ respectively, which can be found in Algorithms \ref{prot:FDNoise} and \ref{prot:FDPerturb} respectively. Here we assume that $\Pi^{(D,L)}$ is a secure protocol that samples a value from $[-L,L]\cap 2^{-p}\mathbb{Z}$ following the distribution $g^{(D,L)},$ which will be discussed in more detail in Section \ref{sec:sampling-gDL}. We also denote by $\Pi^{(\mathtt{GE})},$ the secure protocol that, given an input $[x],$ returns $[0]$ if $x<0$ and $[1]$ otherwise. We note that the security of the proposed protocols is straightforward given the security of the sub-protocols used and hence omitted from our discussion.
\begin{algorithm}[htbp] 
\caption{$([i],[y]) \leftarrow \Pi^{(D)}.\mathtt{Noise}()$} \label{alg:noise}
\label{prot:FDNoise}
\begin{algorithmic}[1]
\REQUIRE Let $E,L,p,\sigma$ be as previously defined. 
\ENSURE The functionality outputs $([i],[y])$ where $(i,y)=\mathcal{F}^{(D)}.\mathtt{Noise}().$
\STATE The parties jointly calculate $[i]\leftarrow \Pi^{(\mathtt{BerSample})}(p)$ where $p=1-\frac{2\cdot 2^p E e^{-\frac{L}{\sigma}}}{\lambda^{(D)}};$
\STATE The parties jointly calculate $[a]\leftarrow \Pi^{(\mathtt{Uni})}(2\cdot 2^pE);$
\STATE The parties jointly calculate $[b]\leftarrow \Pi^{(D,L)}();$
\STATE The parties jointly calculate and return $([i],[a]+[i]\cdot([b]-[a]));$
\end{algorithmic}
\end{algorithm}

\begin{algorithm}[htbp]
\caption{$[z] \leftarrow \Pi^{(D)}.\mathtt{Perturb}([x],[i],[y])$} \label{alg:perturb}
\label{prot:FDPerturb}
\begin{algorithmic}[1]
\REQUIRE Let $x\in [-E,E]\cap 2^{-p}\mathbb{Z}$ be the value we want to perturb and $([i],[y])$ is the output of $\Pi^{(D)}.\mathtt{Noise}()$ 
\ENSURE The functionality outputs $[z]$ such that $z$ follows the distribution $f^{(D)}$
\STATE The parties jointly calculate $[b]\leftarrow \Pi^{(\mathtt{GE})}\left([y]-2^p(E+[x])\right));$
\STATE The parties jointly calculate $[z_0]\leftarrow (2[b]-1)L-E + 2^{-p}([y]+[b]);$
\STATE The parties jointly calculate $[z_1]\leftarrow [x]+[y];$
\STATE The parties jointly calculate and return $[z]\leftarrow[z_0]+[i]\cdot([z_1]-[z_0]);$
\end{algorithmic}
\end{algorithm}

\subsection{2-PC Perturbation of $f^{(C)}$}
We note that $f^{(C)}$ can be considered in a very similar way as $f^{(D)}$ with the following differences. Firstly, the uniform distribution is still for $2\cdot 2^p E$ possible values despite for a slightly different sampling space $S\triangleq ([-L-E,x-L)\cup [x+L,L+E))\cap 2^{-p}\mathbb{Z}$). Furthermore, we require a third functionality $\mathcal{F}^{(I,C)}(p,L)$ which outputs a random element of $T\triangleq [-L,L)\cap 2^{-p}\mathbb{Z}$ where for each possible value $y,$ it is sampled with probability $g^{(C,L)}.$ To realize the perturbation mechanism with output following $f^{(C)},$ we again follow the two-step approach of sampling and perturbation as previously discussed. The specification of $\mathcal{F}^{(C)}.\mathtt{Noise}, \mathcal{F}^{(C)}.\mathtt{Perturb}, \Pi^{(C)}.\mathtt{Noise},$ and $\Pi^{(C)}.\mathtt{Perturb}$ can be found in Algorithms \ref{func:FCNoise}, \ref{func:FCPerturb}, \ref{prot:FCNoise}, and \ref{prot:FCPerturb} respectively. Here we assume that $\Pi^{(C,L)}$ is a secure protocol that samples a value from $[-L,L)\cap 2^{-p}\mathbb{Z}$ following the distribution $g^{(C,L)},$ which will be discussed in more detail in Section \ref{sec:sampling-gCL}. 

\begin{algorithm}[htbp]
\caption{$(i,y) \leftarrow \mathcal{F}^{(C)}.\mathtt{Noise}()$}
\label{func:FCNoise}
\begin{algorithmic}[1]
\REQUIRE Let $E,L,p,\sigma$ be as previously defined. 
\ENSURE The functionality outputs $(i,y)$ where $i=0$ if $y$ is the output of $\mathcal{F}^{(O)}(2\cdot 2^pE)$ and $i=1$ if $y$ is the output of $\mathcal{F}^{(I,C)}(p,L).$ 
\STATE Sample $i\in \{0,1\}$ following a Bernoulli distribution with success probability $1-\frac{2\cdot E e^{-\frac{L}{\sigma}}}{\lambda^{(C)}};$
\IF{$i=0$}
\STATE Sample $a\in\{0,\cdots, 2\cdot 2^pE-1\}$ uniformly at random;
\STATE Return $(i,a);$
\ELSE
\STATE Sample $b\in [-L,L)\cap 2^{-p}\mathbb{Z}$ following the distribution $g^{(C,L)};$
\STATE Return $(i,b);$
\ENDIF
\end{algorithmic}
\end{algorithm}

\begin{algorithm}[htbp]
\caption{$z \leftarrow \mathcal{F}^{(C)}.\mathtt{Perturb}(x,i,y)$}
\label{func:FCPerturb}
\begin{algorithmic}[1]
\REQUIRE Let $x\in [-E,E]\cap 2^{-p}\mathbb{Z}$ be the value we want to perturb and $(i,y)$ is the output of $\mathcal{F}^{(C)}.\mathtt{Noise}()$ 
\ENSURE The functionality outputs $z$ following the distribution $f^{(C)}$
\IF{$i==0$}
\IF{$y\leq 2^p(E+x)-1$}
\STATE Define $z\leftarrow -L-E + y2^{-p};$
\ELSE
\STATE Define $z\leftarrow L-E + 2^{-p}y;$
\ENDIF
\ELSE
\STATE Define $z\leftarrow x+y;$
\ENDIF
\STATE Return $z;$
\end{algorithmic}
\end{algorithm}

\begin{algorithm}[htbp]
\caption{$([i],[y]) \leftarrow \Pi^{(D)}.\mathtt{Noise}()$}
\label{prot:FCNoise}
\begin{algorithmic}[1]
\REQUIRE Let $E,L,p,\sigma$ be as previously defined. 
\ENSURE The functionality outputs $([i],[y])$ where $(i,y)=\mathcal{F}^{(C)}.\mathtt{Noise}().$
\STATE The parties jointly calculate $[i]\leftarrow \Pi^{(\mathtt{BerSample})}(p)$ where $p=1-\frac{2\cdot E e^{-\frac{L}{\sigma}}}{\lambda^{(C)}};$
\STATE The parties jointly calculate $[a]\leftarrow \Pi^{(\mathtt{Uni})}(2\cdot 2^pE);$
\STATE The parties jointly calculate $[b]\leftarrow \Pi^{(C,L)}();$
\STATE The parties jointly calculate and return $([i],[a]+[i]\cdot([b]-[a]));$
\end{algorithmic}
\end{algorithm}

\begin{algorithm}[htbp]
\caption{$[z] \leftarrow \Pi^{(C)}.\mathtt{Perturb}([x],[i],[y])$}
\label{prot:FCPerturb}
\begin{algorithmic}[1]
\REQUIRE Let $x\in [-E,E]\cap 2^{-p}\mathbb{Z}$ be the value we want to perturb and $([i],[y])$ is the output of $\Pi^{(C)}.\mathtt{Noise}()$ 
\ENSURE The functionality outputs $[z]$ such that $z$ follows the distribution $f^{(C)}$
\STATE The parties jointly calculate $[b]\leftarrow \Pi^{(\mathtt{GE})}\left([y]-2^p(E+[x])\right));$
\STATE The parties jointly calculate $[z_0]\leftarrow (2[b]-1)L-E + 2^{-p}[y];$
\STATE The parties jointly calculate $[z_1]\leftarrow [x]+[y];$
\STATE The parties jointly calculate and return  $[z]\leftarrow[z_0]+[i]\cdot([z_1]-[z_0]);$
\end{algorithmic}
\end{algorithm}

\subsection{Protocol $\Pi^{(D,L)}$: Sampling from $g^{(D,L)}$ }\label{sec:sampling-gDL}

We achieve sampling from $g^{(D,L)}$ by transforming a geometric distribution into a discrete Laplace distribution. As defined in Definition~\ref{def:SepTool}, a centred discrete Laplace distribution has a probability mass function of
\[
    g^{(D,S)}(y)= \frac{1}{\lambda^{(D,S)}} e^{-\frac{|y|}{\sigma}},
\]
where $\lambda^{(D,S)}\triangleq \sum_{y\in S} e^{-\frac{|y|}{\sigma}}$ and $S\triangleq [-L,L]\cap  2^{-p}\mathbb{Z}$. To sample from $g^{(D,S)}(y)$, we firstly sample from a scaled version of $g^{(D,S)}$ over the support $2^pS$ and then convert these samples to the support $S$. Let $ y' = 2^{p}y$ where $y'\in S'=[-2^pL,2^pL]\cap\mathbb{Z}.$ We get an equivalent distribution for $g^{(D,S)}(y)$ as 

\[
g^{(D,S')}(y) = \frac{1}{\lambda^{(D,S')}}e^{-\frac{|y'|}{\sigma'}} 
\]
for $\lambda^{(D,S')}=\sum_{x\in S'}e^{-\frac{|x|}{\sigma'}}$ and $\sigma'=2^p\sigma$. For simplicity, we define this truncated discrete Laplace distribution  $\mathcal{L}_{S'}^{(t)}$ as 
\begin{equation*}
    f_{\mathcal{L}}(y;t) = c_0\cdot e^{-|y|/t} \text{~for~} y\in S'.
\end{equation*}
for some normalization constant $c_0$ and constant $t$.  To sample from $\mathcal{L}_{S'}^{(t)}$, we employ the method proposed in \cite{WYF23,canonne2020discrete}.

We define a truncated Geometric distribution $\mathcal{G}^{(t)}_{\kappa,\rho}$ with parameters $\kappa\in \mathbb{Z}_{>0}$ and $\rho\in(0,1)$ and sample space $[0,2^\kappa)\cap \mathbb{Z}.$ Its probability mass function is
\[
f_{\mathcal{G}}(x;\rho)=c_1(1-\rho){\rho}^x ~\text{for~}x\in [0,2^{\kappa})\cap\mathbb{Z}
\]
for some normalization constant $c_1$.

Let $x_{\kappa-1}\cdots x_0$ be the binary expansion of $x$. It is observed in \cite{DKM06} that generating a sample $x$ from $\mathcal{G}^{(t)}_{\kappa,\rho}$ is equivalent to independently Bernoulli sampling every binary digit of $x$. More specifically, for $x\in[0,2^{\kappa})\cap\mathbb{Z}$, one can get the $i$ th bit of $x$ according to a Bernoulli distribution $\mathcal{B}(\rho^{2^i}/(1+\rho^{2^i}))$ for $i=0,1,\cdots,\kappa -1$ ($i=0$ for the least significant bit).

Let $L=2^{\lambda}$ for some non-negative integer $\lambda$.
By setting $\kappa = p+\lambda$ and $\rho=e^{-1/t}$, the distribution of $\mathcal{G}^{(t)}_{\kappa,\rho}$ can viewed as a ``folded" version of the truncated and discrete Laplace $\mathcal{L}_{S'}^{(t)}$ except for a tweak of the probability distribution at $x=0$. To overcome the tweak, we employ the method proposed in \cite{WYF23} to firstly sample over $(-2^{\kappa},-1]\cap[1,2^{\kappa})$ using geometric sampling of $\mathcal{G}^{(t)}_{\kappa,\rho}$ prior to getting the desired $\mathcal{L}_{S'}^{(t)}$ over $S'$. The procedure proceeds as follows (Algorithm~\ref{prot:DLapSampling}).
\begin{enumerate}
    \item Firstly, we invoke a Bernoulli sampler $\mathcal{B}(\frac{1}{\lambda^{(D,S')}})$. If we get an sample of $1$, we return $0$; otherwise, we proceed as below.
    \item Secondly, if the Bernoulli sampler returns $0$, we generate a sample $x'$ from $\mathcal{G}^{(t)}_{\kappa,\rho}$ with $\kappa = p+ \lambda$ using the bitwise method in \cite{DKM06}. Let $x=x'+1$.
    \item Thirdly, we invoke the other Bernoulli sampler $\mathcal{B}(0.5)$ and return $x$ if it generate $1$; otherwise, we return $-x$.
\end{enumerate}

\begin{algorithm}[htbp]
\caption{$[z] \leftarrow \Pi^{(D,L)}(t,c_0)$}
\label{prot:DLapSampling}
\begin{algorithmic}[1]
\REQUIRE The support of $\mathcal{L}_{S'}^{(t)}$ is $x\in [-2^{p}L,2^{p}L]\cap \mathbb{Z}$. $\rho = e^{-1/t}$;$\kappa = \lceil \log 2^pL-1 \rceil$. The parties have access to a Bernoulli protocol $\Pi^{\mathcal{B}}(\rho)$.
\ENSURE The functionality outputs $[z]$ such that $z$ follows the distribution $\mathcal{L}_{S'}^{(t)}$.
\STATE The parties jointly invoke $[b]\leftarrow \Pi^{\mathcal{B}}(\rho)$ for $\rho = c_0$;
\FOR{$i = 0,\cdots,\kappa-1$}
\STATE The parties jointly invoke $[z_i] \leftarrow \Pi^{\mathcal{B}}(\rho^{2^i}/(1+\rho^{2^i}))$;
\ENDFOR
\STATE The parties jointly invoke $[d]\leftarrow \Pi^{\mathcal{B}}(1/2)$, and computes $[s]=2[d]-1$
\STATE The parties jointly compute $[z]=[s]\cdot \left(\sum_i2^i[z_i]+1\right)$.
\STATE The parties jointly calculate and return  $[z]=(1-[b])\cdot[z]\cdot 2^{-p}$.
\end{algorithmic}
\end{algorithm} 

We prove it in the following Proposition that the samples yielded as above follows the truncated and discrete Laplace distribution $\mathcal{L}_{S'}^{(t)}$.

\begin{prop} The samples yielded as above follow the truncated discrete Laplace $\mathcal{L}_{S'}^{(t)}$ over $S'$.\label{pro:samplingLaplace}
\end{prop}
\begin{proof}
    The complete proof can be found in the supplementary material due to page limits.
\end{proof}
\subsection{Protocol $\Pi^{(C,L)}()$: Sampling from $g^{(C,L)}$}\label{sec:sampling-gCL}

Let $S=[-2^{-P}L, 2^{-P}L)\cap\mathbb{Z}$ be the support of the target distribution of $g^{(C,L)}$. Instead of sampling from $g^{(C,L)}$ directly, we utilize the samples from $g^{(D,L)}$ to give an approximate sampling of $g^{(C,L)}$. More specifically,
 the approximate $g^{(C,L)}$ is derived by firstly sampling from $g^{(D,L)}$ over a finer support (e.g. $[-L, L)\cap2^{-(p+\gamma)}\mathbb{Z}$) by Algorithm~\ref{prot:DLapSampling}. Then, given a sample $x$ of $g^{(D,L)}$, we round it down to the closest lattice point on $2^{-p}\mathbb{Z}$. We denote by $\lfloor x \rfloor_{np}$ the \textit{Nearest Plane Algorithm} to solve the \textit{closest vector problem} (CPV). Given a query point $x$ on a finer lattice, the CVP of $x$ on a coarse lattice $2^{-p}\mathbb{Z}$  \cite{Babai1986OnLL} is given by
 \[
 \lfloor x \rfloor_{np} = \lfloor x/2^{-p} \rfloor \cdot 2^{-p}.
 \]
 The finer the fine lattice is, the better derived distribute can approximate $g^{(C,L)}$.

\section{Implementation}\label{sec:imple}
We implemented the truncated discrete Laplace perturbation by the secure 2PC library ABY \cite{demmler2015aby}. The benchmarking was conducted in a controlled laboratory environment on a Dell Precision 7856 workstation equipped with CPU of AMD Ryzen PRO @ 3.8GHz and RAM of 128GB. All experiments were done on a single thread.

\begin{figure}
    \centering
    \begin{subfigure}[b]{0.24\textwidth}
        \centering
        \includegraphics[width=\linewidth]{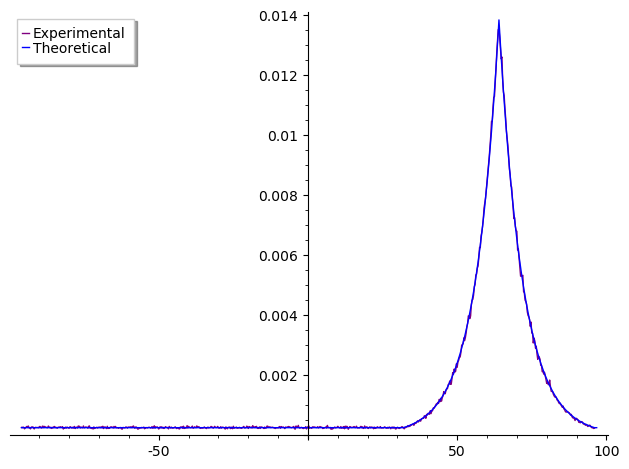}
        \caption{$x=E$}
        \label{fig:sub1}
    \end{subfigure}
    \hfill
    \begin{subfigure}[b]{0.24\textwidth}
        \centering
        \includegraphics[width=\linewidth]{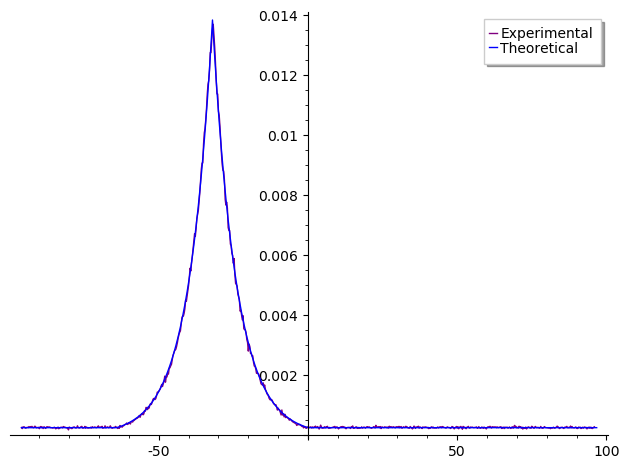}
        \caption{$x=-E/2$}
        \label{fig:sub2}
    \end{subfigure}
    \caption{Distribution of Truncated Discrete Laplace for $E=64$, $L=32$, $\sigma=8$, and precision $p=2$ : Theoretical vs. Experimental}
    \label{fig:impl}
\end{figure}

To validate the correctness of the implementation, in Fig.~\ref{fig:impl}, we depict the ideal truncated discrete Laplace distribution as well as an empirical one obtained from our implementation, where the input value is $x$, $E=64$, $L=32$, $\sigma=8$, and precision $p=2$. The empirical probability mass functions in Fig.~\ref{fig:impl} are derived out of 500,000 samples for each distribution. As shown in the figure, the ideal distribution closely matches the empirical distribution produced by our implementation. We also provide in Table~\ref{tab:accuracy} the empirical results of mean value and MSE across different parameters with the $\epsilon$-DP guarantee as in Proposition \ref{prop:TruncDisLapPriv}, which closely align with the theoretical results derived by Proposition \ref{prop:TruncDisLapEst}. Additionally, we evaluate the complexity of the implementation for the selected parameters in Table~\ref{tab:performance}.

In general, comparison analysis between our method and the existing discrete Gaussian perturbation in their application in distributed scenario with the help of MPC is not straightforward due to the different assumptions and settings between them. Here, we provide a general observation that can be made from our experiment. In this work, we compare our proposed design with the state-of-the-art discrete Gaussian distribution protocol proposed in \cite{WYF23}. Intuitively, the protocol they have proposed first samples from Laplace distribution, which can then be transformed to a discrete Gaussian distribution. We note that our realization of our proposed method uses similar technique as those used in the Laplace distribution sampling step done in \cite{WYF23}. We note that since we are aiming for Laplace distribution instead of Gaussian distribution, we may omit the rejection-and-sampling step which is required in \cite{WYF23}.

To given a concrete example, in their work to achieve $(\epsilon,\delta)-$ DP with  $\epsilon = 1.3$ and $\delta=2^{-\lambda}$ for security parameter $\lambda=64$, the discrete Gaussian requires $\sigma\approx 10$ defined over the support $[-128, 128]\cap \mathbb{Z}$ \footnote{The relationship between $\sigma, \delta, \epsilon$ can be found in Theorem 7 in \cite{canonne2020discrete} and the support $[-N, N]\cap \mathbb{Z}$ is determined by Theorem 4.4 in \cite{WYF23}}. Note that due to the acceptance rate, the size of circuit is influenced by the number of samples they generated in batch and the number of AND gates to generate $n$ samples scales as $O((n+\lambda)\psi \log (\psi \sigma))$ where $\psi = \lambda + \log(n)$. It is indicated in \cite{WYF23} that there are $23.5\times 10^6$ AND gates in the circuit to draw $n=4096$ samples in batch from this discrete Gaussian distribution. That is approximately 15625 AND gates per sample (estimated by the asymptotic complexity). To design truncated discrete Laplace perturbation of the same privacy budget $\epsilon=1.3$, and the same sampling space but zero failure probability, it suffices to instantiate the truncated discrete Laplace distribution with $\sigma=49.2$ and $E=L=64$. The circuit requires 14537 AND gates, which is comparable to the one in \cite{WYF23}. As explained in Section \ref{sec:sampling}, the sampling and perturbing of our DP mechanism are independent. To enhance running time, we can shift the sampling step offline while keeping the perturbing step online.


\begin{table*}[]
    \centering
    \renewcommand{\arraystretch}{1.2}
    \begin{tabular}{|c|c|c|c|c|c|c|}
    \hline
        $\sigma = 8, E=64, L= 32$  & \multicolumn{2}{c|}{$X=0$} & \multicolumn{2}{c|}{$X=-E/2$} & \multicolumn{2}{c|}{$X=E$}  \\ \Xhline{2.5\arrayrulewidth} 
        $p=0$ & experiment  & theoretical &  experiment  & theoretical & experiment  & theoretical  \\ \hline
       Mean Value & 0.12 & 0.00 & -25.04 & -25.75 & 50.92 & 51.49 \\ \hline
       MSE        & 678.32 & 670.66 & 864.36 & 870.75 & 1388.37 & 1471.04 \\    \Xhline{2\arrayrulewidth} 
        $p=2$ & experiment  & theoretical &  experiment  & theoretical & experiment  & theoretical  \\ \hline
       Mean Value & -0.03 & 0.00 & -25.59 & -25.76 & 51.22 & 51.52 \\ \hline
       MSE        & 677.41 & 664.86 & 840.96 & 864.54 &  1331.72& 1463.58 \\ \hline   
    \end{tabular}
    \caption{Accuracy evaluation under $\epsilon$-DP guarantee: theoretical predictions vs. experimental results.}
    \label{tab:accuracy}
\end{table*}

\begin{table*}[]
    \centering
    \renewcommand{\arraystretch}{1.2}
    \begin{tabular}{|p{3cm}|c|c|c|c|}
    \hline
       $\sigma = 8$, $E=64$, $L= 32$  & runtime (ms) & throughput (KiB) & Circuit Max Depth &  No. AND gates \\ \Xhline{2.5\arrayrulewidth}
       $p=0$ & 21.6  & 623.1 & 209 & 14397 \\ \hline
        $p=2$ & 26.6  & 798.8 & 213 & 19781\\ \hline
    \end{tabular}
    \caption{Complexity of the implementation for Algorithm \ref{alg:noise} and \ref{alg:perturb} (security parameter 128 bits).}
    \label{tab:performance}
\end{table*}

\section{Conclusion}
In this work, we proposed two designs of discrete bounded perturbation mechanism that are Laplace-inspired and have zero privacy failure probability. Such mechanisms have then been analyzed in terms of their privacy and utility.

In order to show its compatibility with MPC and other MPC-based solutions, we have also proposed a secure distributed realization of the perturbation mechanism for $2$-party computation. Our MPC realization is shown to be as competitive as the state-of-the-art discrete Gaussian perturbation mechanism in MPC regarding the complexity of the circuit. Additionally, our distributed perturbation mechanism can be divided to sampling phase and perturbing phase. The former can be shifted to offline to enhance runtime efficiency.


\section*{Acknowledgments}
This research is supported by the National Research Foundation, Singapore under its Strategic Capability Research Centres Funding Initiative. Any opinions, findings and conclusions or recommendations expressed in this material are those of the author(s) and do not reflect the views of National Research Foundation, Singapore.

%

\bibliographystyle{ieeetr}
\bibliography{reference}

\newpage
\onecolumn
\appendix

\subsection{Proof of Proposition IV.1}
\begin{proof}
    For simplicity of notation, let $\lambda=\lambda^{(\mathtt{Lap})}_{L,E,\sigma}= \int_{-E-L}^{E+L}e^{-\frac{\min(|y-x|,L)}{\sigma}}dy.$ Then

    \begin{eqnarray*}
        \lambda&=& \int_{-E-L}^{x-L} e^{-\frac{L}{\sigma}}dy + \int_{x+L}^{E+L}e^{-\frac{L}{\sigma}}dy +\int_{x-L}^x e^{-\frac{x-y}{\sigma}}dy+\int_x^{x+L}e^{-\frac{y-x}{\sigma}}dy\\
                &=&e^{-\frac{L}{\sigma}}(((x-L)-(-E-L)) +((E+L)-(x+L))) + \sigma \left(\left.e^{-\frac{x-y}{\sigma}}\right|_{x-L}^x - \left.e^{-\frac{y-x}{\sigma}}\right|_{x}^{x+L}\right)\\
                &=& 2Ee^{-\frac{L}{\sigma}} + \sigma\left(\left(1-e^{-\frac{L}{\sigma}}\right)-\left(e^{-\frac{L}{\sigma}}-1\right)\right)= 2\left((E-\sigma)e^{-\frac{L}{\sigma}} + \sigma\right)
    \end{eqnarray*}  
    completing the proof
\end{proof}
\subsection{Proof of Proposition IV.3}
\begin{proof}
    Let $x\in [-E,E]$ and $\mathbb{E}=\mathbb{E}(\mathcal{M}(x))=\int_{-E-L}^{E+L} yf_{x,\sigma}^{(\mathtt{Lap})}(y) dy.$ Then
    \begin{eqnarray*}
        \lambda^{(\mathtt{Lap})}_{L,E,\sigma} \mathbb{E}&=& \int_{-E-L}^{x-L} ye^{-\frac{L}{\sigma}}dy+ \int_{x+L}^{E+L} ye^{-\frac{L}{\sigma}}dy + \int_{x-L}^x ye^{-\frac{x-y}{\sigma}}dy +\int_{x}^{x+L} ye^{-\frac{y-x}{\sigma}}dy \\
        &=& \frac{e^{-\frac{L}{\sigma}}}{2}\left[(x-L)^2-(-E-L)^2 + (E+L)^2-(x+L)^2\right]\\
        &&+ \int_0^L (x-r)e^{-\frac{r}{\sigma}}dr +\int_0^L (x+r)e^{-\frac{r}{\sigma}}dr\\
        &=& -2xLe^{-\frac{L}{\sigma}} - 2x\sigma \left(\left.e^{-\frac{r}{\sigma}}\right|_{0}^L\right)= x\left(2\left[\sigma(1-e^{-\frac{L}{\sigma}})-Le^{-\frac{L}{\sigma}}\right]\right)= 2x\sigma [1-(1+\epsilon)e^{-\epsilon}]
    \end{eqnarray*}
    This implies that 
    $\mathbb{E}=x\left(\frac{1-(1+\epsilon)e^{-\epsilon}}{1-\left(1-\frac{E}{\sigma}\right)e^{-\epsilon}}\right).$
    Now we consider $\xi=\mathbb{E}((y-x)^2) = \mathbb{E}(y^2) - 2x\mathbb{E}(y) + x^2 = \mathbb{E}(y^2) + x^2\left(1-2\left(\frac{1-(1+\epsilon)e^{-\epsilon}}{1-\left(1-\frac{E}{\sigma}\right)e^{-\epsilon}}\right)\right).$
    
     Let $\mathbb{E}_2=\mathbb{E}[y^2].$ Then, assuming that $\sigma = \frac{L}{\epsilon},$ letting $E=k\sigma$ for some positive real $k,$ we have

    \begin{eqnarray*}
       \lambda^{(\mathrm{Lap})}_{L,E,\sigma} \mathbb{E}_2&=&\int_{-E-L}^{x-L} y^2e^{-\frac{L}{\sigma}}dy + \int_{x+L}^{E+L} y^2e^{-\frac{L}{\sigma}}dy + \int_{x-L}^x y^2e^{-\frac{x-y}{\sigma}}dy +\int_{x}^{x+L} y^2e^{-\frac{y-x}{\sigma}}dy\\
        &=& \frac{e^{-\frac{L}{\sigma}}}{3}\left[(x-L)^3-(-E-L)^3 +(E+L)^3 - (x+L)^3\right]\\
        &&+\int_0^L(x-r)^2 e^{-\frac{r}{\sigma}}dr + \int_0^L(x+r)^2e^{-\frac{r}{\sigma}}dr\\
        &=& \frac{e^{-\frac{L}{\sigma}}}{3}\left[2E^3+6E^2L+6EL^2 - 6x^2L\right]+2x^2\int_0^Le^{-\frac{r}{\sigma}}dr + 2\int_0^L r^2 e^{-\frac{r}{\sigma}}dr\\
        &=&\frac{e^{-\frac{L}{\sigma}}}{3}\left[2E^3+6E^2L+6EL^2 - 6x^2L\right] + 2\sigma x^2\left(1-e^{-\frac{L}{\sigma}}\right)
        \end{eqnarray*}
        \begin{eqnarray*}
        &&+ 4\sigma^3-2\sigma e^{-\frac{L}{\sigma}}\left(L^2 + 2L\sigma + 2\sigma^2\right)\\
        &=& x^2\left(2\sigma\left(1-e^{-\frac{L}{\sigma}}\right)-2Le^{-\frac{L}{\sigma}}\right)+4\sigma^3\\
        && +e^{-\frac{L}{\sigma}}\left(\frac{2}{3}E^3 + 2E^2L + 2EL^2-2\sigma L^2 - 4L\sigma^2 -4\sigma^3\right)\\
        &=&2\sigma (x^2(1-(1+\epsilon)e^{-\epsilon})+2\sigma^2)\\
        &&+2 \sigma^3e^{-\epsilon}\left(\frac{1}{3}k^3 + \epsilon k^2 + \epsilon^2 k - (\epsilon^2+2\epsilon+2)\right)\\
        \end{eqnarray*}

    \begin{claim}\label{claim:Esigmarel}
        For any $\epsilon>0,$ there exists $k^\ast\in(1,2)$ such that $\frac{1}{3}(k^\ast)^3+\epsilon (k^\ast)^2 + \epsilon^2 k^\ast-(\epsilon^2+2\epsilon+2)<0.$ 
    \end{claim}
    \begin{proof}
        Let $f(k) = \frac{1}{3}k^3 + \epsilon k^2 + \epsilon^2 k -(\epsilon^2 + 2\epsilon +2).$ It is easy to see that $f$ is a continuous function and $f(1)=-\epsilon -\frac{5}{3}<0.$ On the other hand, $f(2) = \epsilon^2+2\epsilon +\frac{2}{3}>0.$ Hence there must exist $a\in (1,2)$ such that $f(a)=0.$ Lastly, we note that $f'(k)=(3k^2-1)+2\epsilon(k-1)$ which is positive for any $k\geq 1.$ Hence for any $k^\ast \in (1,a), f(k^\ast)<0.$
    \end{proof}
    
    Choosing $k=k^\ast$ derived in Claim \ref{claim:Esigmarel}, we have 
    \[\mathbb{E}_2\leq \left(\frac{x^2(1-(1+\epsilon)e^{-\epsilon})+2\sigma^2}{(k-1)e^{-\epsilon}+1}\right).\]
    
    Hence 
    \begin{eqnarray*}
        \xi&\leq&x^2\left(\frac{(1-(1+\epsilon)e^{-\epsilon})+2\sigma^2/{x^2}}{(k-1)e^{-\epsilon}+1}+1-2\left(\frac{1-(1+\epsilon)e^{-\epsilon}}{1-\left(1-\frac{E}{\sigma}\right)e^{-\epsilon}}\right)\right)\\
        &=& \frac{2\sigma^2 + x^2e^{-\epsilon}\left(\epsilon+\frac{E}{\sigma}\right)}{1-\left(1-\frac{E}{\sigma}\right)e^{-\epsilon}}
    \end{eqnarray*}

\end{proof}

\subsection{Proof of Proposition V.4}
\begin{proof}
        Let $x\in \mathcal{A}_{p,E}$ and $\mathbb{E}=\mathbb{E}(\mathcal{M}(x)) = \sum_{y\in \mathcal{A}_{p,L+E}} yf_{x,\sigma}^{(\mathtt{DLap})}.$ For simplicity of notation, let $\lambda = \lambda^{(\mathtt{DLap})}_{L,E,\sigma,p}.$ Then $\lambda \mathbb{E} = \sum_{i=0}^4\Delta^{(1)}_i$ where $\Delta^{(1)}_i = \sum_{y\in T^{(x)}_i} ye^{-\frac{\min(|y-x|,L)}{\sigma}}.$ It is easy to see that $\Delta^{(1)}_0=x.$ Furthermore, we have
        \begin{itemize}
            \item For $\Delta^{(1)}_1,$ we note that each summand is in the form of $ye^{-\frac{L}{\sigma}}.$ Hence we have
            \begin{eqnarray*}
                \Delta^{(1)}_1&=& e^{-\frac{L}{\sigma}}\left(\sum_{i=1}^{2^p(x+E)}(x-L-i2^{-p})\right)\\
                &=& e^{-\frac{L}{\sigma}}2^p(x+E)(x-L) - e^{-\frac{L}{\sigma}}\frac{(x+E)(2^p(x+E)+1)}{2}\\
            \end{eqnarray*}
            On the other hand, for $\Delta^{(1)}_4,$ we note that each summand is in the form of $ye^{-\frac{L}{\sigma}}.$ Hence we have
            \begin{eqnarray*}
                \Delta^{(1)}_4&=& e^{-\frac{L}{\sigma}}\left(\sum_{i=1}^{2^p(E-x)}(x+L+i2^{-p})\right)\\
                &=& e^{-\frac{L}{\sigma}}2^p(E-x)(x+L) + e^{-\frac{L}{\sigma}}\frac{(E-x)(2^p(E-x)+1)}{2}.
            \end{eqnarray*} 
            So 
            \begin{eqnarray*}
                \Delta^{(1)}_1+\Delta^{(1)}_4 &=& 2e^{-\frac{L}{\sigma}} 2^p x(E-L) - e^{-\frac{L}{\sigma}} (2\cdot 2^p\cdot E\cdot x + x)\\
                &=& -x\left( 2 e^{-\frac{L}{\sigma}} 2^p L + e^{-\frac{L}{\sigma}}\right)
            \end{eqnarray*}
            \item For $\Delta^{(1)}_2,$ we note that each summand is in the form of $ye^{-\frac{x-y}{\sigma}}.$ Hence we have $
                \Delta^{(1)}_2= \sum_{i=1}^{2^pL}(x-i2^{-p})e^{-\frac{i2^{-p}}{\sigma}}.$
                On the other hand, for $\Delta^{(1)}_3,$ we note that each summand is in the form of $ye^{-\frac{y-x}{\sigma}}.$ Hence we have $
                \Delta^{(1)}_2= \sum_{i=1}^{2^pL}(x+i2^{-p})e^{-\frac{i2^{-p}}{\sigma}}.$ Hence
                \begin{eqnarray*}
                    \Delta_2^{(1)} + \Delta_3^{(1)}&=& 2x\sum_{i=1}^{2^pL} e^{-\frac{i2^{-p}}{\sigma}}\\
                    &=& 2xe^{-\frac{2^{-p}}{\sigma}}\left(\frac{e^{-\frac{L}{\sigma}}-1}{e^{-\frac{2^{-p}}{\sigma}}-1}\right)
                \end{eqnarray*}
        \end{itemize}
        This implies that 
        \begin{eqnarray*}
            \lambda \mathbb{E}&=& x\left(-e^{-\frac{L}{\sigma}} - 2e^{-\frac{L}{\sigma}} 2^p L + 2\left(\frac{1-e^{-\frac{L}{\sigma}}}{e^{\frac{2^{-p}}{\sigma}}-1}\right) \right)\\
            &=& x\left((1-e^{-\frac{L}{\sigma}})\frac{e^{\frac{2^{-p}}{\sigma}}+1}{e^{\frac{2^{-p}}{\sigma}}-1}- \left(1+2\cdot2^pLe^{-\frac{L}{\sigma}}\right)\right)\\
            &=&x\left((1-e^{-\frac{L}{\sigma}})\frac{e^{\frac{2^{-p}}{\sigma}}+1}{e^{\frac{2^{-p}}{\sigma}}-1}- \left(1+2\cdot2^pLe^{-\frac{L}{\sigma}} `\right)\right)
        \end{eqnarray*}
        and
        \[\mathbb{E}=x\left(\frac{(1-e^{-\frac{L}{\sigma}})\frac{e^{\frac{2^{-p}}{\sigma}}+1}{e^{\frac{2^{-p}}{\sigma}}-1}- \left(1+2\cdot2^pLe^{-\frac{L}{\sigma}}\right)}
        {\left(1-e^{-\frac{L}{\sigma}}\right)\left(\frac{e^{\frac{2^{-p}}{\sigma}}+1}{e^{\frac{2^{-p}}{\sigma}}-1}\right) + e^{-\frac{L}{\sigma}}(1+2\cdot 2^p E)}\right).\]
        Let $\kappa$ be such that $\mathbb{E}=\frac{x}{\kappa}.$ Now we consider $\xi = \mathbb{E}((y-x)^2)=\mathbb{E}_2 +x^2 \left(1-\frac{2}{\kappa}\right)$ where $\mathbb{E}_2\triangleq \mathbb{E}(y^2).$ 
        We again let $\lambda\mathbb{E}_2 = \sum_{i=0}^4 \Delta^{(2)}_i$ where $\Delta^{(2)}_i=\sum_{y\in T_i^{(x)}} y^2 e^{-\frac{\min(|y-x|,L)}{\sigma}}.$ Then $\Delta_0^{(2)}=x^2.$ Furthermore, we have
        \begin{itemize}
            \item  For $\Delta^{(2)}_1,$ we note that each summand is in the form of $ye^{-\frac{L}{\sigma}}.$ Hence we have
            \begin{eqnarray*}
                \Delta^{(2)}_1&=& e^{-\frac{L}{\sigma}}\left(\sum_{i=1}^{2^p(x+E)}(x-L-i2^{-p})^2\right)\\
                &=& e^{-\frac{L}{\sigma}}\left(2^p(x+E)(x-L)^2  -(x-L)(x+E)(2^p(x+E)+1)\right.\\
                &&\left.+ \frac{1}{6} (x+E)(x+E+2^{-p})(2\cdot 2^p(x+E)+1)\right)\\
            \end{eqnarray*}
            On the other hand, for $\Delta^{(2)}_4,$ we note that each summand is in the form of $ye^{-\frac{L}{\sigma}}.$ Hence we have
            \begin{eqnarray*}
                \Delta^{(2)}_4&=& e^{-\frac{L}{\sigma}}\left(\sum_{i=1}^{2^p(E-x)}(x+L+i2^{-p})^2\right)\\
                &=& e^{-\frac{L}{\sigma}}\left(2^p(E-x)(x+L)^2 + (x+L) (E-x)(2^p(E-x)+1)\right.\\
                &&\left.+\frac{1}{6} (E-x)(E-x+2^{-p})(2\cdot 2^p(E-x)+1)\right).
            \end{eqnarray*} 

            So 
            \begin{eqnarray*}
                \Delta^{(2)}_1+\Delta^{(2)}_4 &=& e^{-\frac{L}{\sigma}}\left(2\cdot 2^p\left(\frac{E^3}{3}+L^2E+LE^2 - x^2L\right)+ \left(E^2+2LE-x^2\right)+\frac{2^{-p}}{3}E\right) 
            \end{eqnarray*}
            \item For $\Delta^{(2)}_2,$ we note that each summand is in the form of $ye^{-\frac{x-y}{\sigma}}.$ Hence we have $\Delta^{(2)}_2= \sum_{i=1}^{2^pL}(x-i2^{-p})^2e^{-\frac{i2^{-p}}{\sigma}}.$
                On the other hand, for $\Delta^{(2)}_3,$ we note that each summand is in the form of $ye^{-\frac{y-x}{\sigma}}.$ Hence we have $
                \Delta^{(1)}_2= \sum_{i=1}^{2^pL}(x+i2^{-p})^2e^{-\frac{i2^{-p}}{\sigma}}.$ Hence
                \begin{eqnarray*}
                    \Delta_2^{(1)} + \Delta_3^{(1)}&=& 2\sum_{i=1}^{2^pL} (x^2 + i^2 2^{-2p})e^{-\frac{i2^{-p}}{\sigma}}\\
                    &=& 2x^2\left(\frac{1-e^{-\frac{L}{\sigma}}}{e^{\frac{2^{-p}}{\sigma}}-1}\right) + 2\cdot 2^{-2p}\sum_{i=1}^{2^pL} i^2 e^{-\frac{i2^{-p}}{\sigma}}.
                \end{eqnarray*}
                We discuss the second term of the sum above in the following claim. We note that the argument below has been used in various works, for instance in the analysis of geometric distribution, and it is provided below for completeness.
                \begin{claim}\label{claim:secondmomentgeom}
                    Let $p,L,$ and $\sigma$ be as defined in Proposition V.4. Then
                    \begin{enumerate}
                        \item \[\sum_{i=1}^{2^pL} i e^{-\frac{i2^{-p}}{\sigma}}=\frac{(1-e^{-\frac{L}{\sigma}})-2^pL e^{-\frac{L}{\sigma}}}{e^{\frac{2^{-p}}{\sigma}}-1} + \frac{1-e^{-\frac{L}{\sigma}}}{(e^{\frac{2^{-p}}{\sigma}}-1)^2}\]
                        and
                        \item 
                        \begin{eqnarray*}
                        \sum_{i=1}^{2^pL} i^2 e^{-\frac{i2^{-p}}{\sigma}}&=& \left(1-e^{-\frac{L}{\sigma}}\right)\left[\frac{e^{2\frac{2^{-p}}{\sigma}}+e^{\frac{2^{-p}}{\sigma}}}{(e^{\frac{2^{-p}}{\sigma}}-1)^3}\right]\\
                        &&-e^{-\frac{L}{\sigma}}\left[2\cdot 2^pL\frac{e^{\frac{2^{-p}}{\sigma}}}{(e^{\frac{2^{-p}}{\sigma}}-1)^2} + 2^{2p}L^2\frac{1}{e^{\frac{2^{-p}}{\sigma}}-1}\right]
                        \end{eqnarray*}
                    \end{enumerate}
                \end{claim}
                \begin{proof}
                    For a positive integer $n,$ let $f:(0,1)\rightarrow \mathbb{R}$ be the function such that for any $x\in (0,1), f(x)=\sum_{i=1}^{n} x^i= x\frac{1-x^n}{1-x}.$ It is easy to see that $f$ is continuous and twice differentiable at any $x\in(0,1).$ Hence, we have $f'(x) = \frac{1-(n+1)x^n}{1-x} + \frac{x(1-x^n)}{(1-x)^2}.$ On the other hand, we have $f'(x) = \sum_{i=1}^{n} ix^{i-1}.$
                    This implies that $\sum_{i=1}^{n} ix^i = xf'(x) = \frac{x-(n+1)x^{n+1}}{1-x} + \frac{x^2(1-x^n)}{(1-x)^2}$ and setting $n=2^pL$ and $x=e^{-\frac{2^{-p}}{\sigma}}\in (0,1),$ we obtain the first part of the claim.

                    We further note that $f''(x) = \sum_{i=1}^n i(i-1) x^{i-2} = \sum_{i=1}^n i^2 x^{i-2} - \sum_{i=1}^n i x^{i-2}.$ Hence $\sum_{i=1}^n i^2 x^i = x^2 f''(x) +xf'(x).$ On the other hand, we also have 
                    \[
                        f''(x) = -\frac{(n+1)nx^{n-1}}{1-x} + 2\frac{1-(n+1)x^n}{(1-x)^2} + 2\frac{x(1-x^n)}{(1-x)^3}.
                    \]
                    This implies that 
                    \[\sum_{i=1}^n i^2 x^i= \frac{x}{1-x}(1-(n+1)^2 x^n) +\frac{x^2}{(1-x)^2} (3-(2n+3)x^n)+ \frac{2x^3}{(1-x)^3} (1-x^n).\]

                    Substituting $n=2^pL$ and $x=e^{-\frac{2^{-p}}{\sigma}},$ we obtain the second part of the claim.
                \end{proof}
        \end{itemize}
        Combining the discussion above, we obtain
        \begin{eqnarray*}
            \lambda\mathbb{E}_2&=& x^2\left((1-e^{-\frac{L}{\sigma}})\left(1+\frac{2}{e^{\frac{2^{-p}}{\sigma}}-1}\right)-2\cdot 2^p e^{-\frac{L}{\sigma}}L \right)\\
            &&+e^{-\frac{L}{\sigma}}\left(2\cdot 2^p\left(L^2E+LE^2+\frac{E^3}{3}\right) + \left(E^2+2LE-\frac{2L^2}{e^{\frac{2^{-p}}{\sigma}}-1}\right)\right.\\
            &&\left.+2^{-p}\left(\frac{E}{3} - \frac{4Le^{\frac{2^{-p}}{\sigma}}}{(e^{\frac{2^{-p}}{\sigma}}-1)^2}\right)\right)+2\cdot 2^{-2p}(1-e^{-\frac{L}{\sigma}})\frac{e^{\frac{2\cdot 2^{-p}}{\sigma}}+e^{\frac{2^{-p}}{\sigma}}}{(e^{\frac{2^{-p}}{\sigma}}-1)^3}
        \end{eqnarray*}
        and hence
        \begin{eqnarray*}
            \mathbb{E}_2&=& x^2\frac{(1-e^{-\frac{L}{\sigma}})\left(\frac{e^{\frac{2^{-p}}{\sigma}}+1}{e^{\frac{2^{-p}}{\sigma}}-1}\right)-2\cdot 2^p e^{-\frac{L}{\sigma}}L}{\left(1-e^{-\frac{L}{\sigma}}\right)\frac{e^{\frac{2^{-p}}{\sigma}}+1}{e^{\frac{2^{-p}}{\sigma}}-1} + e^{-\frac{L}{\sigma}} (1+2\cdot 2^pE)}\\
            &&+e^{-\frac{L}{\sigma}}\frac{2\cdot 2^p\left(L^2E+LE^2+\frac{E^3}{3}\right) + \left(E^2+2LE-\frac{2L^2}{e^{\frac{2^{-p}}{\sigma}}-1}\right)}{\left(1-e^{-\frac{L}{\sigma}}\right)\frac{e^{\frac{2^{-p}}{\sigma}}+1}{e^{\frac{2^{-p}}{\sigma}}-1} + e^{-\frac{L}{\sigma}} (1+2\cdot 2^pE)}\\
            &&+\frac{e^{-\frac{L}{\sigma}}2^{-p}\left(\frac{E}{3} - \frac{4Le^{\frac{2^{-p}}{\sigma}}}{(e^{\frac{2^{-p}}{\sigma}}-1)^2}\right)}{\left(1-e^{-\frac{L}{\sigma}}\right)\frac{e^{\frac{2^{-p}}{\sigma}}+1}{e^{\frac{2^{-p}}{\sigma}}-1} + e^{-\frac{L}{\sigma}} (1+2\cdot 2^pE)}+ \frac{2\cdot 2^{-2p}(1-e^{-\frac{L}{\sigma}})\frac{e^{\frac{2\cdot 2^{-p}}{\sigma}}+e^{\frac{2^{-p}}{\sigma}}}{(e^{\frac{2^{-p}}{\sigma}}-1)^3}}{\left(1-e^{-\frac{L}{\sigma}}\right)\frac{e^{\frac{2^{-p}}{\sigma}}+1}{e^{\frac{2^{-p}}{\sigma}}-1} + e^{-\frac{L}{\sigma}} (1+2\cdot 2^pE)}.
        \end{eqnarray*}
        This implies that
        \begin{eqnarray*}
            \xi &=& x^2 \left(1-\frac{2}{\kappa}\right) + \mathbb{E}_2
            \end{eqnarray*}
            \begin{eqnarray*}
           &=& x^2\cdot\frac{2+e^{-\frac{L}{\sigma}}(1+2\cdot 2^p(E+L))}{\left(1-e^{-\frac{L}{\sigma}}\right)\frac{1+e^{\frac{2^{-p}}{\sigma}}}{e^{\frac{2^{-p}}{\sigma}}-1}+e^{-\frac{L}{\sigma}}(1+2\cdot 2^p E)}\\
               &&+e^{-\frac{L}{\sigma}}\frac{2\cdot 2^p\left(L^2E+LE^2+\frac{E^3}{3}\right) + \left(E^2+2LE-\frac{2L^2}{e^{\frac{2^{-p}}{\sigma}}-1}\right)}{\left(1-e^{-\frac{L}{\sigma}}\right)\frac{e^{\frac{2^{-p}}{\sigma}}+1}{e^{\frac{2^{-p}}{\sigma}}-1} + e^{-\frac{L}{\sigma}} (1+2\cdot 2^pE)}\\
            &&+\frac{e^{-\frac{L}{\sigma}}2^{-p}\left(\frac{E}{3} - \frac{4Le^{\frac{2^{-p}}{\sigma}}}{(e^{\frac{2^{-p}}{\sigma}}-1)^2}\right)}{\left(1-e^{-\frac{L}{\sigma}}\right)\frac{e^{\frac{2^{-p}}{\sigma}}+1}{e^{\frac{2^{-p}}{\sigma}}-1} + e^{-\frac{L}{\sigma}} (1+2\cdot 2^pE)}+ \frac{2\cdot 2^{-2p}(1-e^{-\frac{L}{\sigma}})\frac{e^{\frac{2\cdot 2^{-p}}{\sigma}}+e^{\frac{2^{-p}}{\sigma}}}{(e^{\frac{2^{-p}}{\sigma}}-1)^3}}{\left(1-e^{-\frac{L}{\sigma}}\right)\frac{e^{\frac{2^{-p}}{\sigma}}+1}{e^{\frac{2^{-p}}{\sigma}}-1} + e^{-\frac{L}{\sigma}} (1+2\cdot 2^pE)}.
        \end{eqnarray*}

        It is then easy to conclude that $\mathrm{Var}(\bar{x})=O(E^2).$ 
        
    \end{proof}

\subsection{Proof of Proposition VI.3}
\begin{proof}
       Let $x,y\in \mathcal{A}_{p,E}$ and $z\in \mathcal{B}_{p,L+E}.$ Let $T_I^{(x)}=[x-L,x+L)\cap 2^{-p}\mathbb{Z}$ while $T_O^{(x)}=\mathcal{B}_{p,L+E}\setminus T_I^{(x)}.$ Note that if $z\in T_I^{(x)}, f_{x,\sigma}^{(\mathtt{CLap})}(z)=\frac{\int_{z}^{z+2^{-p}} e^{-\frac{|r-x|}{\sigma}}dr}{\lambda_{L,E,\sigma}^{(\mathtt{CLap})}}.$ Note that for $x-L<z_1<x<z_2<x+L$ such that $|z_1-x| = |z_2-x|, f_{x,\sigma}^{(\mathtt{CLap})}(z_1) > f_{x,\sigma}^{(\mathtt{CLap})}(z_2).$ Hence for any $z\in T_I^{(x)},$ 
       \[\frac{\sigma e^{-\frac{|x-z|}{\sigma}}e^{-\frac{2^{-p}}{\sigma}}\left(e^{\frac{2^{-p}}{\sigma}}-1\right)}{\lambda_{L,E,\sigma}^{(\mathtt{CLap})}}\leq f_{x,\sigma}^{(\mathtt{CLap})}(z)\leq \frac{\sigma e^{-\frac{|x-z|}{\sigma}}\left(e^{\frac{2^{-p}}{\sigma}}-1\right)}{\lambda_{L,E,\sigma}^{(\mathtt{CLap})}}.\]
       On the other hand, if $z\in T_O^{(x)}, f_{x,\sigma}^{(\mathtt{CLap})}(z) = \frac{2^{-p}e^{-\frac{L}{\sigma}}}{\lambda_{L,E,\sigma}^{(\mathtt{CLap})}}.$ 
       
       Let $\rho=\frac{f_{x,\sigma}^{(\mathtt{CLap})}(z)}{f_{y,\sigma}^{(\mathtt{CLap})}(z)}.$ We consider $\rho$ in four cases depending on the distances between $x,y$, and $z.$
       \begin{enumerate}
           \item It is easy to see that if $z\in T_O^{(x)}$ and $z\in T_O^{(y)}, \rho=1\leq e^{\epsilon|x-y|}$ for any $\epsilon>0.$
           \item Suppose that $z\in T_O^{(x)}$ and $z\in T_I^{(y)}.$ Then
           \[
               \rho\leq \frac{2^{-p}}{\sigma}\frac{1}{1-e^{-\frac{2^{-p}}{\sigma}}}\cdot e^{\frac{|z-y|-L}{\sigma}}.
           \]
           It is easy to see that since $z\in T_O^{(x)}, |z-x|\geq L$ and hence the last term is at most $e^{\frac{|x-y|}{\sigma}}.$ Let $g(a)=\ln\left(\frac{a}{1-e^{-a}}\right).$ Then $\rho\leq g\left(\frac{2^{-p}}{\sigma}\right) e^{g\left(\frac{2^{-p}}{\sigma}\right)+\frac{|x-y|}{\sigma}}.$ It is easy to see that $g(a)\geq 0$ for any $a> 0$ and it is increasing with respect to $a.$ 
            \begin{claim}\label{claim:gub}
                For any $a>0, g(a)\leq \frac{a}{2}.$
            \end{claim}
            \begin{proof}
                Note that proving the claim is equivalent to proving that $e^{\frac{a}{2}}-e^{-\frac{a}{2}}-a\geq 0$ for $a>0.$ Let $h(a)=e^{\frac{a}{2}}-e^{-\frac{a}{2}}-a.$ Then $h'(a) = \frac{1}{2}\left(e^{\frac{a}{2}}+e^{-\frac{a}{2}}\right)-1.$ Now note that for any $b\geq 0$ by the relation of arithmetic and geometric mean, $b+\frac{1}{b}\geq 2.$ Since $e^{\frac{a}{2}}\geq 0$ for any $a>0,$ we have that $h'(a)\geq 0,$ that is, $h$ is an increasing function. This implies that $h(a)\geq h(0)=0,$ proving that $g(a)\leq \frac{a}{2}$ for any $a>0.$
            \end{proof}
            By Claim \ref{claim:gub}, we have $\rho\leq e^{\frac{1}{\sigma}\left(\frac{2^{-p}}{2} + |x-y|\right)}.$ Note that $|x-y|\geq 2^{-p}.$ Hence 
            \[\rho\leq e^{\frac{1}{\sigma}\left(\frac{2^{-p}}{2} + |x-y|\right)}\leq e^{\frac{3}{2\sigma}|x-y|}.\]
            This implies that the mechanism satisfies the $(\epsilon,d)$-DP if $\sigma \geq \frac{2}{\epsilon}\geq \frac{3}{2\epsilon}$ proving the claim for this second case.
            \item Suppose that $z\in T_I^{(x)}$ and $z\in T_O^{(y)}.$ Then
            \[\rho\leq e^{\frac{L-|x-z|}{\sigma}}\cdot\frac{e^{\frac{2^{-p}}{\sigma}}-1}{\frac{2^{-p}}{\sigma}}.\]
            Similar to the second case, consider $g(a)=\ln\left(\frac{e^a-1}{a}\right).$ It can be observed that for any $0<a<1, g(a)\leq a.$ Hence, it also applies when $a=\frac{2^{-p}}{\sigma}.$ We further note that since $z\in T_I^{(x)}, |z-x|>L.$ This implies that $L-|x-z|\leq 0.$ Hence noting that $|x-y|\geq 2^{-p},$ when $\sigma \geq \max\left(\frac{2}{\epsilon},2^{-p}\right)\geq \frac{3}{2\epsilon},$
            \begin{eqnarray*}
                \rho\leq e^{\frac{2^{-p}}{\sigma}}\leq e^{\frac{|x-y|}{\sigma}}\leq e^{\frac{2}{3}\epsilon |x-y|}\leq e^{\epsilon|x-y|},
            \end{eqnarray*}
            proving the claim for case $3.$
            \item Lastly, suppose that $z\in T_I^{(x)}\cap T_I^{(y)}.$ Then by using the same observation of $2^{-p}\leq |x-y|,$ we have
            \[\rho\leq \frac{e^{-\frac{|x-z|}{\sigma}}}{e^{-\frac{2^{-p}}{\sigma}}e^{-\frac{|y-z|}{\sigma}}}\leq e^{\frac{|x-y|+2^{-p}}{\sigma}}\leq e^{\frac{2}{\sigma}|x-y|}\leq e^{\epsilon |x-y|}\]
            proving the claim for the last case.
       \end{enumerate}
         
   \end{proof}

\subsection{Proof of Proposition VI.4}
\begin{proof}
For any $x\in \mathcal{A}_{p,E},$ we partition $\mathcal{B}_{p,L+E}$ to five sets:
\begin{enumerate}
    \item $T_0^{(x)}=\{-L-E,x+L\}.$ We note that $f_{x,\sigma}^{(\mathtt{CLap})}(y) = \frac{2^{-p}e^{-\frac{L}{\sigma}}}{\lambda^{(\mathtt{CLap})}_{L,E,\sigma}}$ for $y\in T_0^{(x)}.$
    \item $T_1^{(x)}=(-L-E,x-L)\cap 2^{-p}\mathbb{Z}=\{x-L-i2^{-p}:i=1,\cdots, 2^p(E+x)-1\}.$ Note that for $y\in T_1^{(x)},f_{x,\sigma}^{(\mathtt{CLap})}(y) = \frac{2^{-p}e^{-\frac{L}{\sigma}}}{\lambda^{(\mathtt{CLap})}_{L,E,\sigma}}.$
    \item $T_2^{(x)} = [x-L,x)\cap 2^{-p}\mathbb{Z}=\{x-i2^{-p}: i=1,\cdots, 2^pL\}.$ Note that for $y\in T_2^{(x)}, f_{x,\sigma}^{(\mathtt{CLap})}(y)=\frac{\sigma e^{-\frac{x-y}{\sigma}}\left(e^{\frac{2^{-p}}{\sigma}}-1\right)}{\lambda^{(\mathtt{CLap})}_{L,E,\sigma}}.$
    \item $T_3^{(x)} = [x,x+L)\cap 2^{-p}\mathbb{Z}=\{x+i2^{-p}:i=0,\cdots, 2^pL-1\}.$ Note that for $y\in T_3^{(x)}, f_{x,\sigma}^{(\mathtt{CLap})}(y)=\frac{\sigma e^{-\frac{y-x}{\sigma}}\left(1-e^{-\frac{2^{-p}}{\sigma}}\right)}{\lambda^{(\mathtt{CLap})}_{L,E,\sigma}}.$ Furthermore, it is easy to verify that for any $i=1,\cdots, 2^pL, f_{x,\sigma}^{(\mathtt{CLap})}(x-i2^{-p})=f_{x,\sigma}^{(\mathtt{CLap})}(x+(i-1)2^{-p}),$ providing a one-to-one correspondence between elements of $T_2^{(x)}$ and $T_3^{(x)}.$ 
    \item $T_4^{(x)}=(x+L,E+L)\cap 2^{-p}\mathbb{Z}=\{x+L+i2^{-p}:i=1,\cdots, 2^p(E-x)-1\}.$ Note that for $y\in T_4^{(x)}, f_{x,\sigma}^{(\mathtt{CLap})}(y)=\frac{2^{-p} e^{-\frac{L}{\sigma}}}{\lambda^{(\mathtt{CLap})}_{L,E,\sigma}}.$
\end{enumerate} 

    Let $\mathbb{E}=\mathbb{E}(\mathcal{M}(x)) = \sum_{y\in \mathcal{B}_{p,L+E}} yf_{x,\sigma}^{(\mathtt{CLap})}.$ For simplicity of notation, let $\lambda = \lambda_{L,E,\sigma,p}^{(\mathtt{CLap})}.$ First we consider $\lambda \mathbb{E}$ and divide the sum to $5$ parts, $\lambda\mathbb{E} = \sum_{j=0}^4 \Delta_j^{(1)}$ where $\Delta_j^{(1)} \sum_{y\in T_j^{(x)}} y\int_y^{y+2^{-p}} e^{-\frac{\min(|r-x|,L)}{\sigma}}dr.$ Then
    \begin{itemize}
        \item When $j=0,$ simple algebraic calculation yields
        \[\Delta_0^{(1)} = (x-E)2^{-p}e^{-\frac{L}{\sigma}}.
        \]
        \item When $j=1,$ we have 
        \[\Delta_1^{(1)} = (x-L)e^{-\frac{L}{\sigma}}\left(E+x-2^{-p}\right) - \frac{1}{2} e^{-\frac{L}{\sigma}}\left(E+x-2^{-p}\right)(E+x).\]
        On the other hand, when $i=4,$ we have
        \[\Delta_4^{(1)} = (x+L)e^{-\frac{L}{\sigma}}\left(E-x-2^{-p}\right) + \frac{1}{2} e^{-\frac{L}{\sigma}}\left(E-x-2^{-p}\right)(E-x).\]
        Hence 
        \begin{eqnarray*}
            \Delta_1^{(1)} + \Delta_4^{(1)} &=& -xe^{-\frac{L}{\sigma}}\left(2L+2^{-p}\right)
        \end{eqnarray*}
        \item For $j=2,3,$ due to the symmetry we have previously observed, we have
        \begin{eqnarray*}
            \Delta_2^{(1)}+\Delta_3^{(1)}&=& (2x-2^{-p}) \sigma \left(e^{\frac{2^{-p}}{\sigma}}-1\right)\sum_{i=1}^{2^pL} e^{-\frac{i2^{-p}}{\sigma}}\\
            &=&(2x-2^{-p}) \sigma (1-e^{-\frac{L}{\sigma}}).
        \end{eqnarray*}
    \end{itemize}
    Combining these, we obtain that 
    \[
        \lambda \mathbb{E}= x\left(2\sigma(1-e^{-\frac{L}{\sigma}})-2Le^{-\frac{L}{\sigma}}\right) -\left(2^{-p}Ee^{-\frac{L}{\sigma}} + 2^{-p}\sigma (1-e^{-\frac{L}{\sigma}})\right)
    \]
    which implies that
    \[\mathbb{E} = x\left(\frac{\sigma(1-e^{-\frac{L}{\sigma}})-e^{-\frac{L}{\sigma}}L}{\sigma(1-e^{-\frac{L}{\sigma}})+e^{-\frac{L}{\sigma}}E}\right)-\frac{2^{-p}}{2}.\]
    Let $\kappa = \frac{\sigma(1-e^{-\frac{L}{\sigma}})+e^{-\frac{L}{\sigma}}E}{\sigma(1-e^{-\frac{L}{\sigma}})-e^{-\frac{L}{\sigma}}L}.$ 

    Now we consider $\xi.$ It is easy to see that $\xi=\mathbb{E}_2 + 2^{-p}x + x^2\left(1-\frac{2}{\kappa}\right)$ where $\mathbb{E}_2=\mathbb{E}[y^2].$ Let $\lambda \mathbb{E}_2 = \sum_{j=0}^4 \Delta_j^{(2)}$ where $\Delta_j^{(2)}=\sum_{y\in T_i^{(x)}} y^2 \int_{y}^{y+2^{-p}}e^{-\frac{\min(|r-x|,L)}{\sigma}}dr.$ Then

    \begin{itemize}
        \item It is easy to see that
        \[\Delta_0^{(2)}=(2L^2 + E^2 + x^2 + 2LE+2xL)2^{-p}e^{-\frac{L}{\sigma}}.\]
        \item For $j=1,$ we have
        \begin{eqnarray*}
            \frac{1}{e^{-\frac{L}{\sigma}}}\Delta_1^{(2)}&=&\sum_{i=1}^{2^p(x+E)-1}(x-L-i2^{-p})^2 2^{-p}\\
            &=& (x-L)^2(x+E-2^{-p})-(x-L)(x+E)(x+E-2^{-p}) \\
            &&+ \frac{1}{6}(x+E-2^{-p})(x+E)(2(x+E)-2^{-p})\\
            &=& (x-L)^2(x+E)-(x-L)(x+E)^2 +\frac{1}{3}(x+E)^3\\
            &&- 2^{-p}\left[(x-L)^2-(x-L)(x+E)+\frac{1}{2}(x+E)^2\right]\\
            &&+2^{-2p}\cdot \frac{1}{6}(x+E)\\
            &=& L^2x-Lx^2 + L^2E + E^2L + \frac{x^3}{3} + \frac{E^3}{3}\\
            &&- 2^{-p}\left[L^2 - Lx + LE + \frac{1}{2}x^2 + \frac{1}{2}E^2\right] + 2^{-2p}\frac{1}{6}(x+E).
        \end{eqnarray*}
        On the other hand, for $j=4,$ we have
        \begin{eqnarray*}
            \frac{1}{e^{-\frac{L}{\sigma}}}\Delta_4^{(2)}&=&\sum_{i=1}^{2^p(E-x)-1}(x+L+i2^{-p})^2 2^{-p}\\
            &=& (x+L)^2(E-x-2^{-p})+(x+L)(E-x)(E-x-2^{-p}) \\
            &&+ \frac{1}{6}(E-x-2^{-p})(E-x)(2(E-x)-2^{-p})\\
            &=& (x+L)^2(E-x)+(x+L)(E-x)^2 +\frac{1}{3}(E-x)^3\\
            &&- 2^{-p}\left[(x+L)^2+(x+L)(E-x)+\frac{1}{2}(E-x)^2\right]\\
            &&+2^{-2p}\cdot \frac{1}{6}(E-x)\\
            &=& -L^2x-Lx^2 + L^2E + E^2L - \frac{x^3}{3} + \frac{E^3}{3}\\
            &&- 2^{-p}\left[L^2 + Lx + LE + \frac{1}{2}x^2 + \frac{1}{2}E^2\right] + 2^{-2p}\frac{1}{6}(E-x).
        \end{eqnarray*}
        This implies that 
        \begin{eqnarray*}
            \frac{\Delta_1^{(2)}+\Delta_4^{(2)}}{e^{-\frac{L}{\sigma}}} &=& 2L^2E+2E^2L-2Lx^2+\frac{2E^3}{3}\\
            &&- 2^{-p}\left(2L^2+2LE+x^2+E^2\right) + 2^{-2p}\frac{E}{3}.
        \end{eqnarray*}
        Furthermore, combining with $\Delta_0^{(2)},$ we have
        \begin{eqnarray*}
            \Delta_0^{(2)}+\Delta_1^{(2)}+\Delta_4^{(2)} &=&e^{-\frac{L}{\sigma}}\left(2L^2E+2E^2L-2Lx^2+\frac{2E^3}{3} + 2\cdot2^{-p}xL + \frac{2^{-2p}E}{3}\right)\\
            &=&-2Le^{-\frac{L}{\sigma}} x^2 + 2\cdot 2^{-p}Le^{-\frac{L}{\sigma}} x \\
            &&+ e^{-\frac{L}{\sigma}}\left(2L^2E+2E^2L+\frac{2E^3}{3} + \frac{2^{-2p}E}{3}\right).
        \end{eqnarray*}
        \item Next we consider the cases when $j=2$ and $j=3.$ Recall that for $i=1,\cdots, 2^pL,$ we have 
        \[\int_{x-i\cdot 2^{-p}}^{x-(i-1).2^{-p}} e^{-\frac{x-r}{\sigma}}dr = \int_{x+(i-1)\cdot 2^{-p}}^{x+i2^{-p}} e^{-\frac{r-x}{\sigma}}dr.\]
        Hence, following the same argument as Claim \ref{claim:secondmomentgeom},
        \begin{eqnarray*}
            \Delta_2^{(2)}+\Delta_3^{(2)} &=& \sum_{i=1}^{2^pL}\left((x-i\cdot 2^{-p})^2+(x+(i-1)\cdot 2^{-p})^2\right)\sigma e^{-\frac{i2^{-p}}{\sigma}}\left(e^{\frac{2^{-p}}{\sigma}}-1\right)\\
            &=&(2x^2-2x2^{-p}+2^{-2p})\sigma (e^{\frac{2^{-p}}{\sigma}}-1)\sum_{i=1}^{2^pL} e^{-\frac{i2^{-p}}{\sigma}}\\
            &&+ 2\cdot 2^{-2p}\sigma (e^{\frac{2^{-p}}{\sigma}}-1)\cdot e^{-\frac{2\cdot 2^{-p}}{\sigma}}\sum_{i=1}^{2^pL}i(i-1) e^{-\frac{(i-2)2^{-p}}{\sigma}}\\
            &=& \sigma (2x^2 -2x2^{-p}+2^{-2p}) (1-e^{-\frac{L}{\sigma}})\\
            &&+ 2\cdot 2^{-2p}\sigma (e^{\frac{2^{-p}}{\sigma}}-1)\left(-\frac{(2^pL+1)2^pLe^{-\frac{L}{\sigma}}}{e^{\frac{2^{-p}}{\sigma}}-1} + 2\frac{1-(2^pL+1)e^{-\frac{L}{\sigma}}}{(e^{\frac{2^{-p}}{\sigma}}-1)^2}\right.\\
            &&\left.+ 2\frac{1-e^{-\frac{L}{\sigma}}}{(e^{\frac{2^{-p}}{\sigma}}-1)^3}\right)\\
            &=&\sigma (2x^2 -2x2^{-p}+2^{-2p}) (1-e^{-\frac{L}{\sigma}})\\
            &&+2\sigma\left(-(L+2^{-p})Le^{-\frac{L}{\sigma}} -2\cdot2^{-p}L\frac{e^{-\frac{L}{\sigma}}}{e^{\frac{2^{-p}}{\sigma}}-1} +2\cdot 2^{-2p}\frac{1-e^{-\frac{L}{\sigma}}}{e^{\frac{2^{-p}}{\sigma}}-1}\right.\\
            &&\left.+ 2\cdot 2^{-2p}\frac{1-e^{-\frac{L}{\sigma}}}{(e^{\frac{2^{-p}}{\sigma}}-1)^2}\right)\\
            &=& 2\sigma(1-e^{-\frac{L}{\sigma}})x^2 - 2\sigma 2^{-p}(1-e^{-\frac{L}{\sigma}})x \\
            &&- \sigma\left(L^2e^{-\frac{L}{\sigma}} + 2^{-p}Le^{-\frac{L}{\sigma}}\frac{e^{\frac{2^{-p}}{\sigma}}+1}{e^{\frac{2^{-p}}{\sigma}}-1} - 2\cdot 2^{-2p}(1-e^{-\frac{L}{\sigma}})\cdot\frac{e^{\frac{2^{-p}}{\sigma}}}{(e^{\frac{2^{-p}}{\sigma}}-1)^2}\right).
        \end{eqnarray*}
        
    \end{itemize}
    Hence 
        \begin{eqnarray*}
            \lambda \mathbb{E}_2 &=& x^2 \left(2\sigma (1-e^{-\frac{L}{\sigma}})-2Le^{-\frac{L}{\sigma}}\right) - 2^{-p}x \left(2\sigma (1-e^{-\frac{L}{\sigma}})-2Le^{-\frac{L}{\sigma}}\right)\\
            &&+ e^{-\frac{L}{\sigma}} \left(2L^2E + 2E^2L + \frac{2E^3}{3} - \sigma L^2\right) - 2^{-p}L\sigma e^{-\frac{L}{\sigma}}\frac{e^{\frac{2^{-p}}{\sigma}}+1}{e^{\frac{2^{-p}}{\sigma}}-1}\\
            &&+ 2\cdot 2^{-2p} \left[\sigma(1-e^{-\frac{L}{\sigma}})\left(\frac{e^{\frac{2^{-p}}{\sigma}}}{(e^{\frac{2^{-p}}{\sigma}}-1)^2}\right)+\frac{Ee^{-\frac{L}{\sigma}}}{6}\right]
        \end{eqnarray*}
        which implies 
        \begin{eqnarray*} 
            \mathbb{E}_2 &=& \frac{x^2}{\kappa} - 2^{-p}\frac{x}{\kappa}+ e^{-\frac{L}{\sigma}}\frac{2L^2E+2E^2L+\frac{2E^3}{3} - \sigma L^2}{2(\sigma(1-e^{-\frac{L}{\sigma}})+Ee^{-\frac{L}{\sigma}})}\\
            &&-2^{-p} \sigma e^{-\frac{L}{\sigma}}\frac{e^{\frac{2^{-p}}{\sigma}}+1}{e^{\frac{2^{-p}}{\sigma}}-1}\frac{L}{2(\sigma(1-e^{-\frac{L}{\sigma}})+Ee^{-\frac{L}{\sigma}})}\\
            &&+2^{-2p}\frac{\sigma(1-e^{-\frac{L}{\sigma}})\left(\frac{e^{\frac{2^{-p}}{\sigma}}}{(e^{\frac{2^{-p}}{\sigma}}-1)^2}\right)+\frac{Ee^{-\frac{L}{\sigma}}}{6}}{\sigma(1-e^{-\frac{L}{\sigma}})+Ee^{-\frac{L}{\sigma}}}.
        \end{eqnarray*}
        Recall that
        \[
            \xi = \mathbb{E}_2 + 2^{-p}x + x^2\left(1-\frac{2}{\kappa}\right).
        \]
        Hence
        \begin{eqnarray*}
            \xi&=& \left(x^2+2^{-p}x\right)\frac{e^{-\frac{L}{\sigma}}(E+L)}{\sigma\left(1-e^{-\frac{L}{\sigma}}\right)+Ee^{-\frac{L}{\sigma}}}+ e^{-\frac{L}{\sigma}}\frac{2L^2E+2E^2L+\frac{2E^3}{3} - \sigma L^2}{2(\sigma(1-e^{-\frac{L}{\sigma}})+Ee^{-\frac{L}{\sigma}})}\\
            &&-2^{-p} \sigma e^{-\frac{L}{\sigma}}\frac{e^{\frac{2^{-p}}{\sigma}}+1}{e^{\frac{2^{-p}}{\sigma}}-1}\frac{L}{2(\sigma(1-e^{-\frac{L}{\sigma}})+Ee^{-\frac{L}{\sigma}})}\\
            &&+2^{-2p}\frac{\sigma(1-e^{-\frac{L}{\sigma}})\left(\frac{e^{\frac{2^{-p}}{\sigma}}}{(e^{\frac{2^{-p}}{\sigma}}-1)^2}\right)+\frac{Ee^{-\frac{L}{\sigma}}}{6}}{\sigma(1-e^{-\frac{L}{\sigma}})+Ee^{-\frac{L}{\sigma}}}.
        \end{eqnarray*}
    Noting that $L=O(E),$ it is easy to see that
    $\xi= O(E^2)+ O(\sigma).$
    So since $\sigma=O(E^2),$ we have that $\xi=O(E^2),$ proving the claim.
\end{proof}

\subsection{Proof of Proposition VII.1}
   \begin{proof}
   The target distribution is $\mathcal{L}_{S'}^{(t)}$ as 
    \begin{equation*}
        f_{\mathcal{L}}(y;t) = c_0\cdot e^{-|y|/t} \text{~for~} y\in S'.
    \end{equation*}

   In Section VII-D, at the first step of the procedure to convert a geometric distribution into a Laplacian distribution, we get $0$ with probability $c_0$. Otherwise, we proceed with step 2.
 At the second step, a shifted and discrete geometric distribution $\mathcal{G}^{(t)}_{\kappa,\rho}$ over support $S'+1$ is defined as
    \begin{align*}
        f_{\mathcal{G'}}(x'=x+1;p) & = c_1(1-p)p^{(x'-1)}   \stackrel{(a)}{=} c_2 e^{-x'/t}
    \end{align*}
    where $x'\in [1,2^{\kappa}]\cap\mathbb{Z}$, $(a):~p=e^{-1/t} \text{~and~}c_2=c_1(1-e^{-1/t})/e^{-1/t}$. Then, we sample the sign $s$ with a probability of $0.5.$
    At the third step, the final outcome $z$ is derived by
    \begin{equation*}
    z = 
        \begin{cases}
            0    & w.p.~c_0,  \\
            (1-2s)\cdot x'& otherwise.
        \end{cases}
    \end{equation*}
    The probability of $z=v$ is
    \begin{equation*}
        Pr\{z=v\} =
        \begin{cases}
            p_0=f_{\mathcal{L}}(0;t) &~\text{if}~v=0, \\
            (1-p_0)\cdot 1/2 \cdot c_2e^{-|v|/t} &~ \text{if}~v\in[-2^{\kappa},-1]\cap[1,2^{\kappa}]\cap\mathbb{Z}.
        \end{cases}
    \end{equation*}
Let $c_3=(1-p_0)\cdot 1/2 \cdot c_2$. We have $Pr\{z\}=c_3e^{-|z|/t}$ for $z\in S'\backslash 0$; and $Pr\{z=0\}=f_{\mathcal{L}}(0;t)$. Therefore, we can conclude 
\[
Pr\{z\}=f_{\mathcal{L}}(z;t),
\]
 for $z \in S'$.


\end{proof}

\vfill

\end{document}